\documentclass[review]{elsarticle}

\usepackage{lineno,hyperref}
\modulolinenumbers[5]

\journal{Nuclear Instruments and Methods in Physics Research Section A}

\begin{document}

\begin{frontmatter}

\title{Non-linearity effects on the light-output calibration of light charged particles in CsI(Tl) scintillator crystals}

\author[a]{D.~Dell'Aquila\corref{correspondingauthor}}
\ead{dellaqui@msu.edu}
\author[a,b]{S.~Sweany}
\author[a]{K.W.~Brown}
\author[c]{Z.~Chajecki}
\author[a,b]{W.G.~Lynch}
\author[a,b]{F.C.E.~Teh}
\author[a,b]{C.-Y.~Tsang}
\author[a,b]{M.B.~Tsang}
\ead{tsang@nscl.msu.edu}
\author[a,b]{K.~Zhu}
\author[a]{C.~Anderson}
\author[a,b]{A.~Anthony}
\author[d]{S.~Barlini}
\author[a,b]{J.~Barney}
\author[d]{A.~Camaiani}
\author[a]{G.~Jhang}
\author[a,b]{J.~Crosby}
\author[a,b]{J.~Estee}
\author[a,b]{M.~Ghazali}
\author[g]{F.~Guan}
\author[c]{O.~Khanal}
\author[a]{S.~Kodali}
\author[h]{I.~Lombardo}
\author[a,b]{J.~Manfredi}
\author[e,f]{L.~Morelli}
\author[a]{P.~Morfouace}
\author[a]{C.~Niu}
\author[h]{G.~Verde}

\address[a]{National Superconducting Cyclotron Laboratory, Michigan State University, East Lansing, MI 48824, USA}
\address[b]{Department of Physics - Michigan State University, East Lansing, Michigan, USA}
\address[c]{Department of Physics - Western Michigan University, Kalamazoo, Michigan, USA}
\address[d]{Dipartimento di Fisica, Università degli Studi di Firenze and INFN-Sezione di Firenze, 50019 Firenze, Italy}
\address[e]{GANIL, CEA/DSM-CNRS/IN2P3, F-14076 Caen, France}
\address[f]{Dipartimento di Fisica, Università degli Studi di Bologna, 40126 Bologna, Italy}
\address[g]{Tsinghua University, Beijing, China}
\address[h]{INFN-Sezione di Catania, 95123 Catania, Italy}

\cortext[correspondingauthor]{Corresponding author}

\begin{abstract}
The light output produced by light ions ($Z\leq4$) in CsI(Tl) crystals is studied over a wide range of detected energies ($E\leq300$ MeV). Energy-light calibration data sets are obtained with the $10$ cm crystals in the recently upgraded High-Resolution Array (HiRA10).  We use proton recoil data from $^{40,48}$Ca + CH$_2$ at $28$ MeV/u, $56.6$ MeV/u, $39$ MeV/u and $139.8$ MeV/u and data from a dedicated experiment with direct low-energy beams. We also use the punch through points of p, d, and t particles from $^{40,48}$Ca + $^{58,64}$Ni, $^{112,124}$Sn collisions reactions at $139.8$ MeV/u. Non-linearities, arising in particular from Tl doping and light collection effects in the CsI crystals, are found to significantly affect the light output and therefore the calibration of the detector response for light charged particles, especially the hydrogen isotopes. A new empirical parametrization of the hydrogen light output, $L(E,Z=1,A)$, is proposed to account for the observed effects. Results are found to be consistent for all $48$ CsI(Tl) crystals in a cluster of $12$ HiRA10 telescopes.
\end{abstract}


\end{frontmatter}

\nolinenumbers

\section{Introduction}
Thallium-activated caesium iodide (CsI(Tl)) scintillation crystals with photodiode (PD) readout are widely used in nuclear and particle physics experiments. They provide a popular solution to detect and identify, with good energy resolution (usually on the order of a few percent) and excellent identification performance, gamma-rays, light charged particles and intermediate-mass fragments (IMF) produced in nuclear collisions. CsI(Tl) crystals combine several advantageous features including (1) relatively low manufacturing costs, (2) almost non-hygroscopic behaviors, (3) particle-dependent light output that enables particle identification via pulse-shape analysis, and (4) high density ($\rho\approx4.5$ g/cm$^3$). The high-density of the material is a particularly important feature since it allows stopping for highly energetic particles with relatively short detectors, minimizing the sensitivity to surface effects in the light response \cite{Meijer87}. These characteristics make CsI(Tl) crystals particularly well-suited for being used as residual energy stages for silicon-based telescope systems, enabling the identification of fragments via the so-called $\Delta$E-E technique \cite{LeNeindre02}. Many large acceptance detection systems such as INDRA \cite{Pout95}, CHIMERA \cite{Russ15,Lomb10}, NIMROD-ISiS \cite{Wuen09}, GARFIELD \cite{Mor16} or FAZIA \cite{Sal16,Past17}, have been used successfully in detecting and identifying particles in wide energy and mass ranges. Recently, detectors based on modular strip Si-CsI telescopes have been developed to provide high angular resolution in addition to high energy resolution for nuclear structure, particle-particle correlation and nuclear reaction studies. LASSA \cite{Davin01}, MUST2 \cite{Poll05}, FARCOS \cite{Acosta12,Pag16} and HiRA \cite{Wall07} are examples of these modular systems. Such detectors are typically designed to have an optimized response to the detection of lighter particles and to have a high degree of versatility compared to the previous generation of large-acceptance detection systems. The higher stopping power of charged particles in CsI crystals enables a significantly larger dynamic energy range than that achieved with Si-Si modular telescopes (such as OSCAR \cite{Dell18}). However, the energy calibrations are more complicated as the light response of the CsI crystals is not linear and depends on the charge and mass of the charged particles as well as the length of the detector.
The light response of CsI(Tl) scintillators is known to be non-linear for highly ionizing charged particles, due to the quenching phenomenon. Quenching is microscopically caused by an inefficient charge carrier recombination in the presence of high carrier density (eventually enhanced by lattice imperfections due to the interaction of a slowing down fragment) \cite{Par02a}, and, therefore, is linked to the stopping power ($dE/dx$) of the incident radiation. The connection of the scintillation differential-light output to the ion stopping power in inorganic crystals was proposed by Birks in the 1960’s \cite{Birks64}. In addition, the production of energetic $\delta$-electrons affects the light output of inorganic crystals for heavy ions at high energy \cite{Meyer62} making it impossible to describe the scintillation efficiency of CsI crystals uniquely as a function of the radiation stopping power alone. To account for this, the authors of Refs.~\cite{Par02a,Par02b,Lopez18} recently developed the "recombination and nuclear quenching model" (RNQM), that was successfully and systematically applied to produce energy calibrations of heavy ions ($2\leq Z\leq60$) in CsI(Tl), readout by photomultipliers (PMTs), with an accuracy better than $3\%$ for $E/A>5$ MeV and $5$ to $15\%$ at lower energies. For light charged particles with $Z\leq 6$, the approximation $dE/dx\propto-AZ^2/E$ has been used to analytically integrate the differential light output of Ref.~\cite{Birks64} into an expression easier to use \cite{Horn92,Laroc94} and with an explicit dependence on the charge (Z) and mass (A) of the fragment, with an accuracy usually better than $5\%$, resulting in an almost linear trend at low stopping powers. However, significant deviations from the linearity have been observed in Refs.~\cite{Meijer87,Twen90}, and more recently in Ref.~\cite{Aiello96}, in the medium and high energy region of $Z\leq 3$ isotopes, suggesting the existence of possible additional efficiency factors that could play a role in the energy-light calibration of CsI scintillators. This clearly demands more extended investigations of the CsI(Tl) light output to light charged particles.
In this paper, we study the light output of long ($\approx10$ cm) CsI(Tl) crystals incorporated in the upgraded High-Resolution Array (HiRA10) detection system. Our effort focuses particularly on the case of hydrogen isotopes, for which a reliable calibration is particularly needed in view of our recent efforts to explore collective properties of the emission of protons, as well as of other light fragments, produced in heavy ion collisions. Benefiting from the length of the HiRA10 crystals, we exploited high quality data collected in our recent experimental campaign at the National Cyclotron Superconducting Laboratory (NSCL), combined with low energy data from a dedicated experiment at the tandem accelerator of Western Michigan University, to build a consistent CsI calibration data set for hydrogen isotopes (from $1$ MeV up to energies of around $200$ MeV $^1$H and $300$ MeV $^3$H). For particles with $2\leq Z\leq 4$, the detected energy of the particles is up to $250$ MeV. Mainly as a result of the crystal non-uniformities in the Tl doping and in the light collection efficiency, the light output to hydrogen isotopes is found to exhibit a non-linear trend. Based on our observations, we propose a new empirical calibration formula to describe the hydrogen light output with a unique set of parameters, accounting for the observed non-linearity effects. Such effects are found to be negligible within the explored energy range for heavier ($Z\leq2$) isotopes.
The paper is organized as follows: Section 2 describes the experimental procedure used to obtain CsI calibration data sets, Section 3 provides a discussion of the effect of crystal non-uniformities on the light output; finally, results reported in Section 3 are used as the starting point to construct the CsI energy calibration described in Section 4.
\section{Experimental details and calibration data sets}
\subsection{The experimental setup: HiRA10}
\begin{figure}[t]
	\begin{center}
		\includegraphics[scale=0.2]{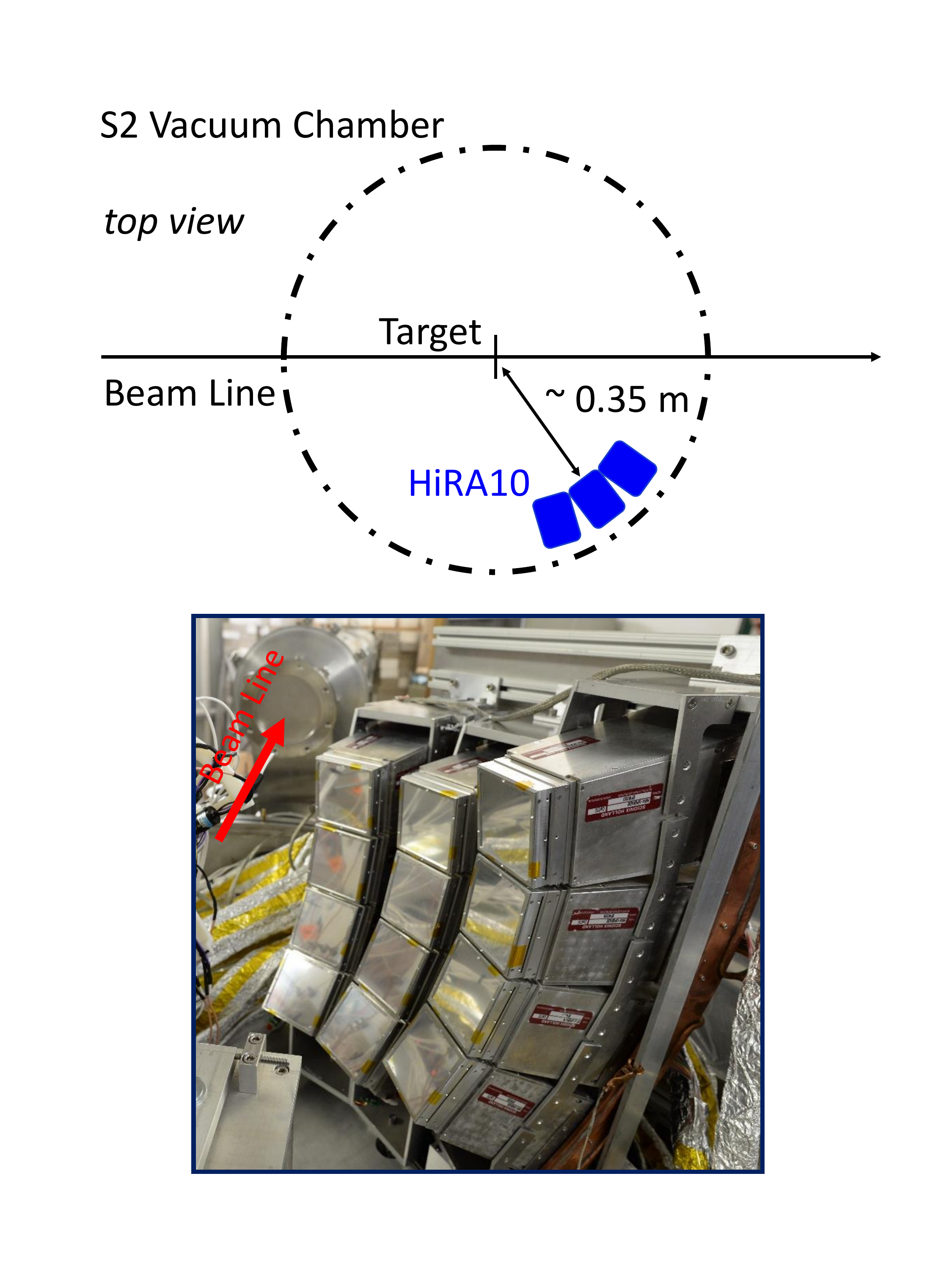}\hfil
		\includegraphics[scale=0.2]{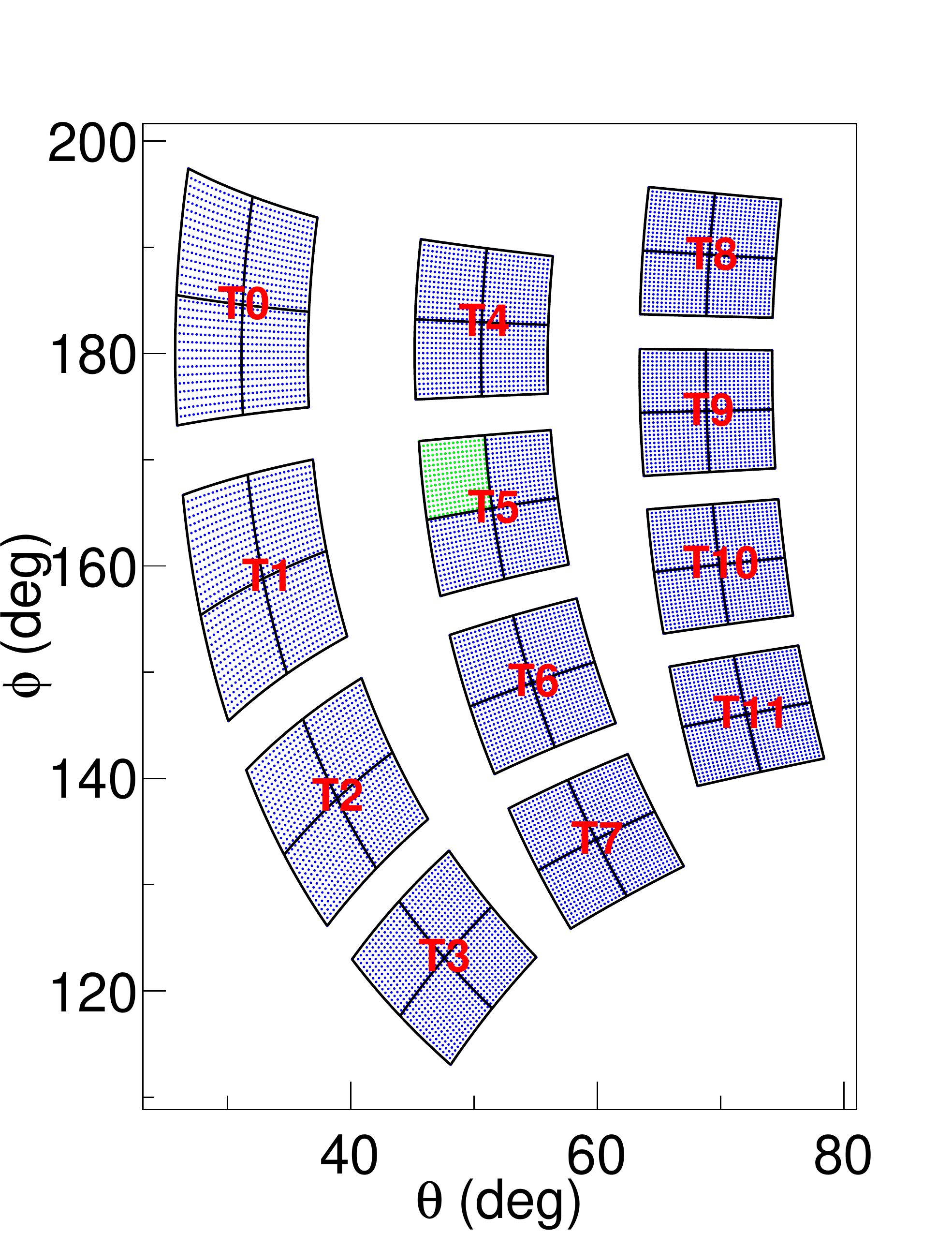}
		\caption{(left panel) The HiRA setup used during the NSCL experiment. Three distinct HiRA towers, each containing 4 vertically arranged telescopes, are installed as shown by the bottom picture. The polar angle region covered by the cluster is visible in the top schematic view. (right panel) The angular range covered by the cluster of 12 HiRA10 telescopes ($48$ CsI crystals) used in the NSCL experimental campaign shown on a ($\theta$,$\phi$) plane. The telescope number in the cluster is indicated by red labels, each blue point represents a DSSSD pixel, and black lines indicate the edges of CsI crystals. Green points indicate the DSSSD pixels corresponding to the crystal used as the typical case for the present paper.}
		\label{figure_01}
	\end{center}
\end{figure}
Experimental data used to calibrate the light output of the HiRA10 CsI(Tl) crystals are obtained by combining several independent data sets: (1) data extracted by using $\Delta$E-E loci and the Ziegler energy loss tables \cite{Zieg85}, (2) proton recoil scattering data, (3) hydrogen isotope punch-through points, and (4) low energy elastic scattering data on bare crystals. Such an extended data set allows for effective constraint on the light output of HiRA10 CsI crystals for $^1$H ($1\leq E\leq198$ MeV), $^2$H ($1\leq E\leq263$ MeV), $^3$H ($10\leq E\leq312$ MeV) and the heavier ions $^{3,4,6}$He, $^{6,7,8}$Li, and $^{7,9}$Be up to around $400$ MeV.  Data sets $1$-$3$ are obtained as a byproduct of our recent experimental campaign at the NSCL at Michigan State University (MSU). ${40,48}$Ca beams were accelerated by means of the NSCL K500 and K1200 coupled cyclotrons with an energy of $139.8$ MeV/u. Degraders were inserted along the beam line to provide two extra energy-degraded beams with energies of $56.6$ MeV/u and $28$ MeV/u. The undegraded beam and the two degraded beams were delivered to the experimental hall by means of the A1900 fragment separator with a precision ranging from $0.2$ to $0.5\%$ energy resolution. The experiments used $^{58,64}$Ni and $^{112,124}$Sn targets. For calibration purposes, a $10$ mg/cm$^2$ thick CH$_2$ target was used to obtain the proton recoil as described below. The position of the beam impinging on the target was monitored during beam tuning by means of a camera installed in the vacuum chamber and a luminescent viewer with a calibrated reference scale. The emitted charged particles from nuclear collisions induced by the Ca beams are detected by $12$ HiRA10 telescopes arranged in $3$ towers in the $53"$ vacuum chamber in the S2 vault, as visible in the top view in the left panel of Figure~\ref{figure_01}. The HiRA10 array covers angles ranging from around $25$ to $75$ degrees in the laboratory frame and an azimuthal angular region shown by the right panel of Figure~\ref{figure_01}. Each HiRA10 telescope consists of a DSSSD (Double-Sided Si Strip Detector) of around $1500$ $\mu$m thickness backed by an array of $4$ closely packed CsI(Tl) scintillator crystals. 
\begin{figure}[t]
	\begin{center}
		\includegraphics[scale=0.4]{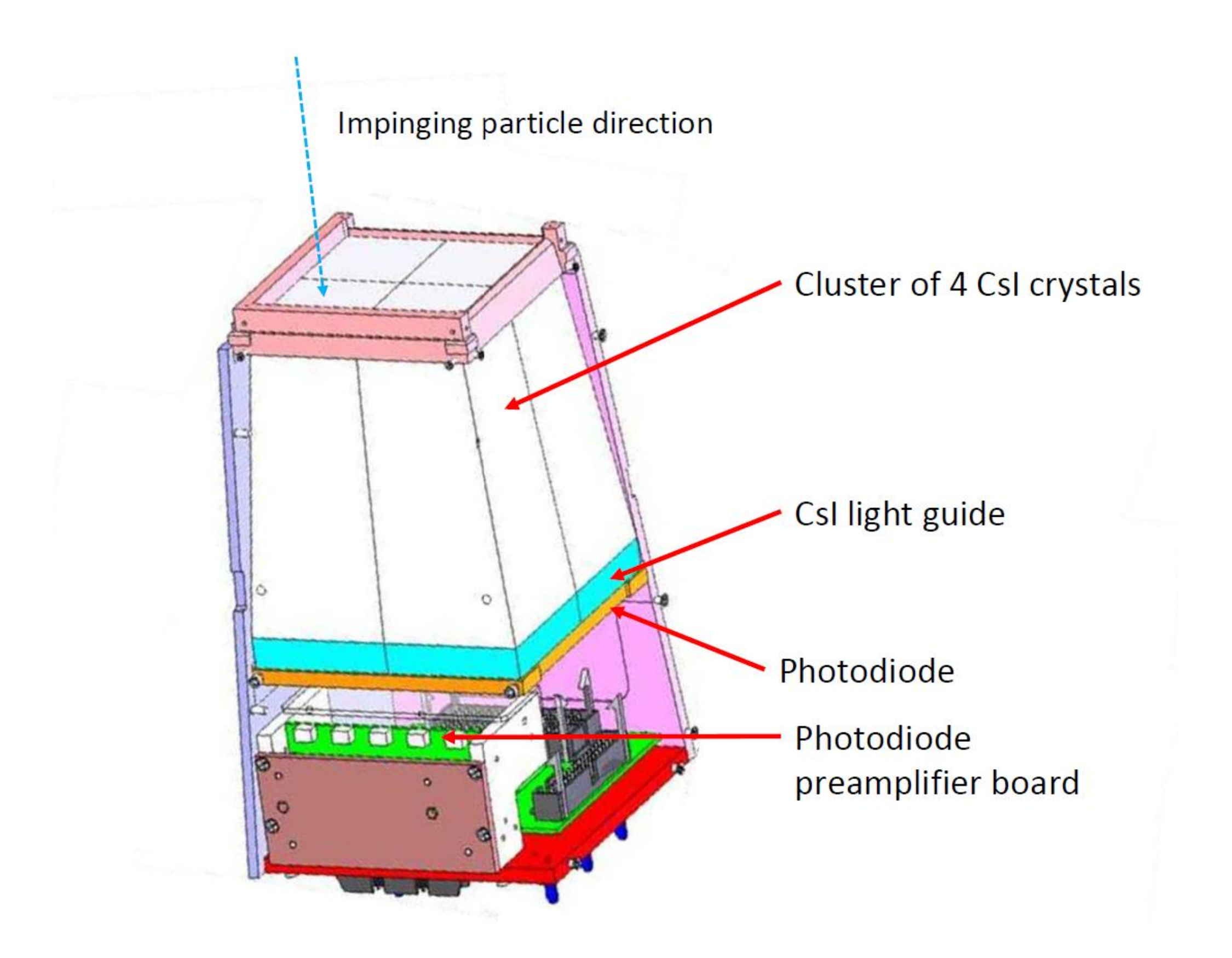}
		\caption{A schematic of the second detection stage of a HiRA10 telescope with container and associated on-board electronics (pre-amplifier board and connections). The dashed arrow indicates the direction of the impinging particles. The first detection stage (DSSSD) is placed at the entrance window of the $4$ CsI crystals (not shown in the figure for clarity reasons).}
		\label{figure_02}
	\end{center}
\end{figure}
Figure~\ref{figure_02} shows a schematic of the assembly of $4$ HiRA10 CsI(Tl) crystals in their container. In the figure, the incident direction of the detected particles is represented by the dashed arrow. The DSSSD detection stage, not shown in the figure for clarity reasons, was installed in front of the entrance face of the crystals. The crystals, manufactured by SCIONIX Scintillation Detectors \cite{Scionix}, are arranged in a $2\times2$ configuration covering approximately an area of $70\times70$ mm$^2$ at their front surface. Each crystal has longitudinal length of $10$ cm, a front surface of around $35\times35$ mm$^2$ and a back surface covering an area of around $45\times45$ mm$^2$, designed for the cluster to be placed at a distance of $35$ cm from the target. The front surface is finished with a fine polish to provide an almost ideal inner reflection of the light produced in the crystal and the best uniformity of the light output even for particles that barely penetrate into the crystal. A thin aluminized Mylar foil ($0.29$ mg/cm$^2$) is used as a wrapping for the entrance widow of each crystal. All other surfaces are machined to an opaque finish (with $200$ grit sandpaper) and coupled with a white reflective wrapping ($0.3$ mm thick) providing an efficient diffuse reflectivity. To account for the thickness of the lateral wrapping, the container holding the $4$ crystals allows a space of around $0.6$ mm in between adjacent crystals. This introduces a thin dead zone for the detection of particles striking a narrow region between crystals, which is taken into account in the analysis by excluding signals produced by particles that hit the first detection stage in a front or a  back strip located in front of a dead region. The back-side surface of each crystal is optically coupled to a square $45 \times 45$ mm$^2$ Plexiglas light guide with a longitudinal length of $1$ cm that provides an optical connection to a $18 \times 18$ mm$^2$ $0.3$ mm thick Hamamatsu pin diode used as the photodiode for the light output conversion. The photodiode is finally coupled to the light guide with flexible optically transparent RTV adhesive. As shown on Figure~\ref{figure_02}, the container hosting the $4$ crystals has an embedded circuit board containing the photodiode preamplifiers and their connections to a flat cable connector located at the back side of the container, which provides all the necessary connections for the DSSSD and CsI detectors. The CsI-thallium doping was chosen to have a concentration of more than $1200$ ppm in order to maximize the light response and obtain good quality energy resolution. The uniformity of the light output over different points of the entrance window and along the crystal length were tested at NSCL. Before fabrication, a $^{241}$Am alpha-source collimated to a spot size of $3$ mm was used to scan the front and back surface of a prototype crystal over $9$ spots located along two diagonals on the entrance window of the crystal. This provides a test of the light output produced by a particle stopping close to the front surface  of the crystal. As specified, non-uniformities are found to be less than $\pm 1.0\%$, testifying the good homogeneity of thallium along sections of the crystal orthogonal to its longitudinal direction. After fabrication of the final crystal, the uniformity of each crystal over the entrance surface was more extensively tested with a $^{241}$Am alpha-source and an $8\times8$ grid configuration. Results are found to be consistent with those obtained with the prototype. To test the longitudinal distribution of thallium inside the crystal, a collimated $^{137}$Cs gamma source was used to laterally scan the finished crystal across different depths. The light response was found to exhibit a linearly decreasing trend as a function of the crystal depth, with overall non-uniformity values not exceeding $4\%$-$5\%$ between entrance and exit windows of the crystal (see Figure~\ref{figure_09}). This is a key point of the present paper and will be discussed in more detail in Section 3. 
\subsection{The energy loss method}
\begin{figure}[t]
	\begin{center}
		\includegraphics[scale=0.5]{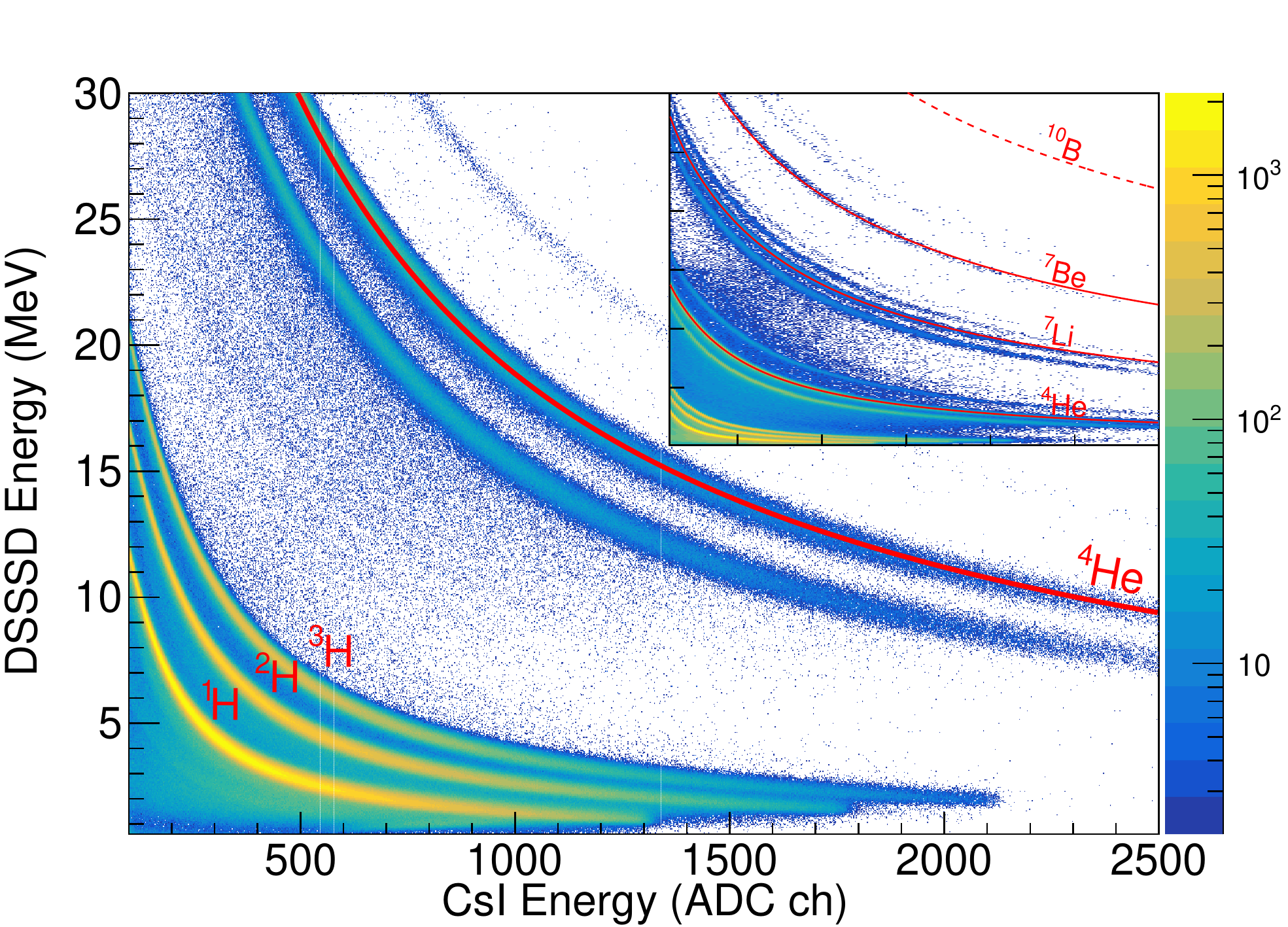}
		\caption{Calibrated DSSSD energy loss ($\Delta E$) versus uncalibrated residual energy in the CsI ($E_{res}$) shown for one HiRA10 CsI crystal, $16$ DSSSD front strips, and $16$ DSSSD back strips for $^{40,48}$Ca collisions at $56.5$ and $138.9$ MeV/u on various targets. Loci corresponding to different identified isotopes are clearly visible. The x-axis is expressed in ADC channels. The top right inset shows an extended plot up to $120$ MeV of energy loss $\Delta E$. Red lines are the result of our simultaneous fit of data shown for a sample of isotopes for clarity reasons (see text for details).}
		\label{figure_03}
	\end{center}
\end{figure}  
If the incident particle is completely stopped in the CsI detection stage, the correlation of the energy deposited in the first detection stage (usually indicated as $\Delta$E) and its residual energy measured by the corresponding CsI can be used to identify the mass and charge of the incident particle (the $\Delta$E-E identification technique). An example is shown in Figure~\ref{figure_03}, which is produced by combining NSCL experiment data for each beam and target combination for one of the HiRA10 crystals and $16$ strips, front and back, of one telescope. For clarity reasons, the figure and the inset show the light ($Z=1$, $Z=2$) and medium ($Z<5$) mass regions, respectively. Higher charge and mass isotopes can also be well identified but are produced with lower statistics in the angular range covered by HiRA10 during the experiment. The plot shows the DSSSD energy calibrated in MeV and the residual energy in the CsI crystal in uncalibrated ADC channels. Silicon energy calibrations have been carefully performed for each front and back strip by using $4$ peaks of a $^{232}$U $\alpha$-source, spanning an effective energy range from $5.41$ MeV to $8.58$ MeV. To correct for the energy loss by the $\alpha$-particles in the thin aluminum layer on the front side of the silicon, we have considered an equivalent silicon dead layer of $0.6$ $\mu$m, as obtained from a previous dedicated investigation \cite{Man18}. DSSSD electronics linearity has been verified by sending a series of calibrated pulses to the preamplifier of each strip. Taking advantage of the high-quality calibrations of the DSSSD, one can extract, for each individual CsI ADC channel $E_{CsI}^{ch}$, the y-axis centroid ($\overline{\Delta E}$) of the calibrated $\Delta$E distribution obtained by gating on the $\Delta$E-E locus of a certain isotope. The kinetic energy of the isotope impinging on the DSSSD layer ($\overline{E}$) can be deduced by a numerical inversion of Ziegler’s energy loss tables \cite{Zieg85} using $\overline{\Delta E}$. The energy impinging on the CsI crystal ($E_{CsI}^{MeV}$) is then calculated by subtracting the energy loss by the ion in the silicon, the dead layer located at its exit face and the Mylar foil used to wrap the CsI entrance face: $E_{CsI}^{MeV}= \overline{E}-\overline{\Delta E}-\Delta E_{dead layer}-\Delta E_{Mylar}$, which corresponds to the energy associated with the initial ADC raw channel $E_{CsI}^{ch}$. A proper implementation of this technique relies on a precise knowledge of the energy deposited in the DSSSD and of its thickness. For this reason, silicon thickness values provided by the manufacturer, and ranging from $1460$ $\mu$m to $1537$ $\mu$m for all the telescopes in the array, have been benchmarked with a dedicated study \cite{Man18}. However, this technique is limited only to the low or middle end of the CsI dynamic range. When the energy of the incident ion is high, a large range of residual CsI energies correspond to roughly the same $\Delta E$. This effect is made worse by the energy straggling experienced by charged particles through the DSSSD stage. To account for this,  we have limited the region of the CsI light output calibrated by this method up to around $60$ MeV for $^1$H, $110$ MeV for $^2$H, $150$ MeV for $^3$H and $200$ MeV for helium isotopes. Because of the more limited statistics recorded for heavier isotopes in the angular range covered by the crystals, $\overline{\Delta E}$ centroids for different $E_{CsI}^{ch}$ values were extracted by an analytical curve obtained by fitting the observed $\Delta$E-E curves with a multi-parametric formula in the case of the lithium and beryllium isotopes \cite{LeNeindre02}. The best fit has been achieved with an individual set of parameters constrained with a simultaneous fit of all visible isotopes in the range $1\leq Z\leq4$. Results of the fitting procedure are found consistent with those obtained by individually fitting $\overline{E}$ centroids in each CsI ADC channel for the higher statistics isotopes ($Z\leq2$). A sample of the obtained curves is shown in Figure~\ref{figure_03} by red lines for some isotopes. A dashed line shows, as an example, the prediction of $^{10}$B, for which the recorded statistics does not allow a firm analysis. Because of the small inaccuracies of our fit at high energies, we have limited the energy range constrained by this method for lithium and beryllium isotopes to $300$ MeV. 
\subsection{Proton-recoil scattering data}
\begin{figure}[t]
	\begin{center}
		\includegraphics[scale=0.35]{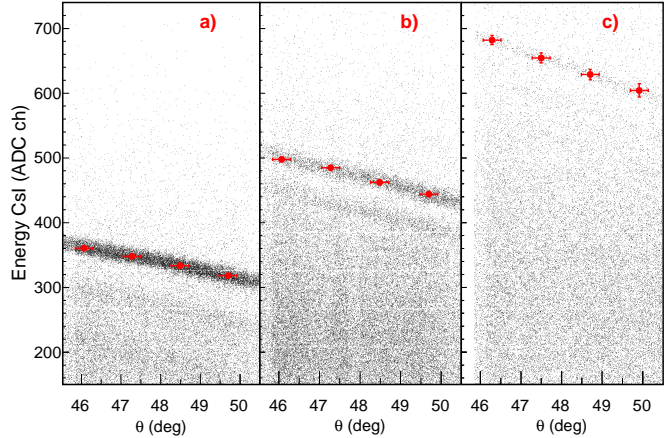}
		\caption{Recoiling proton kinematic lines CsI Energy vs theta for a) $^{48}\textrm{Ca}+^1\textrm{H}$ at $28$ MeV/u, b) $^{40}\textrm{Ca}+^1\textrm{H}$ at $39$ MeV/u and c) $^{48}\textrm{Ca}+^1\textrm{H}$ MeV/u for one HiRA10 crystal. To extract the angle position within the crystal, we used the pixel information provided by the perpendicular crossing of 16 front strips and $16$ back strips, covering the whole surface of the crystal. Red points with error bars represent the calibration points extracted with a fit procedure for some angle slices. Vertical error bars reflect the statistical error on the fit, while horizontal error bars reflect the width of the angular cut chosen and are affected by systematic errors in the beam position and beam spot on the experiment target.}
		\label{figure_04}
	\end{center}
\end{figure} 
Proton-recoil scattering data are extracted from $^{40,48}$Ca + $^1$H elastic scattering measured at incident energies of $28$ MeV/u, $56.6$ MeV/u, $39$ MeV/u and $139.8$ MeV/u on a CH$_2$ target. Recoiling protons from the target are detected in most of the HiRA10 array telescopes, allowing for simultaneous constraint of the proton light output for $48$ crystals in different energy regions depending on the position of the crystal in the array and on the beam energy. The polar angle ($\theta$) formed by the scattered proton direction with the incident-beam direction was calculated by using the position of the proton impinging on the DSSSD detector, known from the coincidence of front and back strips. This relies on a precise knowledge of the absolute position of each DSSSD detector in the array with respect to the collision vertex. The relative position of the reaction target plane, as well as the one of each DSSSD, has been carefully measured with a high-accuracy ROMER Arm instrument \cite{RomerArm}. To extract the absolute DSSSD position with respect to the beam line, a successive laser measurement was performed to fix reference points in the scattering chamber and the detectors to the FRIB/NSCL global reference frame. The angles at which the outgoing protons were detected have been corrected for the geometrical center of the beam spot on the target, as monitored periodically during the experiment for each individual beam using a blank target. Additionally, to account for the uncertainties of the beam position, the shape of the beam cross-section on the target, recorded by the phosphorus viewer at the end of beam tuning procedure, has been used to construct an angular interval of confidence for each detection pixel. The angular resolution is determined with an accuracy better than 0.5 degrees and is majorly limited by the indetermination of the beam position and direction.
Light-energy calibration points are obtained by analyzing the kinematic curves of the elastically recoiling proton for each crystal and for each beam energy. An example for one of the analyzed crystals (covering the angular domain indicated by the green DSSSD pixels of the right panel of Figure~\ref{figure_01}), is shown in Figure~\ref{figure_04}. The three panels show data obtained from 3 different incident energies of the Ca beam: a)$^{48}$Ca @ $28$ MeV/u, b) $^{40}$Ca @ $39$ MeV/u and c) $^{48}$Ca @ $56.5$ MeV/u from the left to the right. Here the x-axis represents the detection polar angle as obtained with the previously described procedure, while the y-axis is the recorded light output by the CsI in ADC channels. Three distinct lines for increasing energies from the left to the right are clearly observed in the plots. They lie on a continuous background due to proton-emitting reactions. In the $39$ MeV/u plot (panel b), an additional line is also visible, corresponding to the inelastic  scattering events leading to the first excited states in the scattered $^{40}$Ca projectile. Due to lower statistics when compared to the ground state elastic events, the inelastic scattering data have been excluded from this analysis. The decreasing statistics in the population of the elastic scattering line as the energy increases is due to the more forward focused kinematics of the scattered projectile. If the projectile is scattered at a smaller angle with respect to the incident direction, the recoiling proton has a more backward peaked distribution, resulting in more statistics for more backward telescopes in the HiRA10. Each well identified elastic scattering line is then divided into $4$ angular bins for each crystal and a Gaussian fit of the corresponding CsI ADC channel distribution is used to extract the light output produced by elastically scattered protons in a certain angular bin. The corresponding energy in MeV is then assigned based on the detection angle ($\theta$) while its uncertainty reflects the angular determination of the selected region. This uncertainty, shown in Figure~\ref{figure_04} as the horizontal error bar of each extracted point, is deduced by combining the error on the size of the angular slice and the angular uncertainty assigned to each pixel in the DSSSD as a result of the beam cross-section on the target. Additionally, calculated energies for each angular bin have been carefully corrected by considering the energy loss by protons in the actual distance travelled in the experimental target and in the SnPb ($18$ $\mu$m) and Mylar ($1.47$ $\mu$m) foils placed at the entrance window of the DSSSD during the experiment. The aluminized mylar foil is part of the telescope housing to form a Faraday cage and to protect the Si surface from dust, light, and pump oil. The SnPb foils stop the $\delta$-electrons abundantly produced in heavy ion collisions. Finally, the energy loss in the silicon front and back dead layer, as well as the aluminized Mylar used to wrap the front face of the crystal, is also taken into account.
\subsection{Particle punch-through calibration points}
\begin{figure}[t]
	\begin{center}
		\includegraphics[scale=0.5]{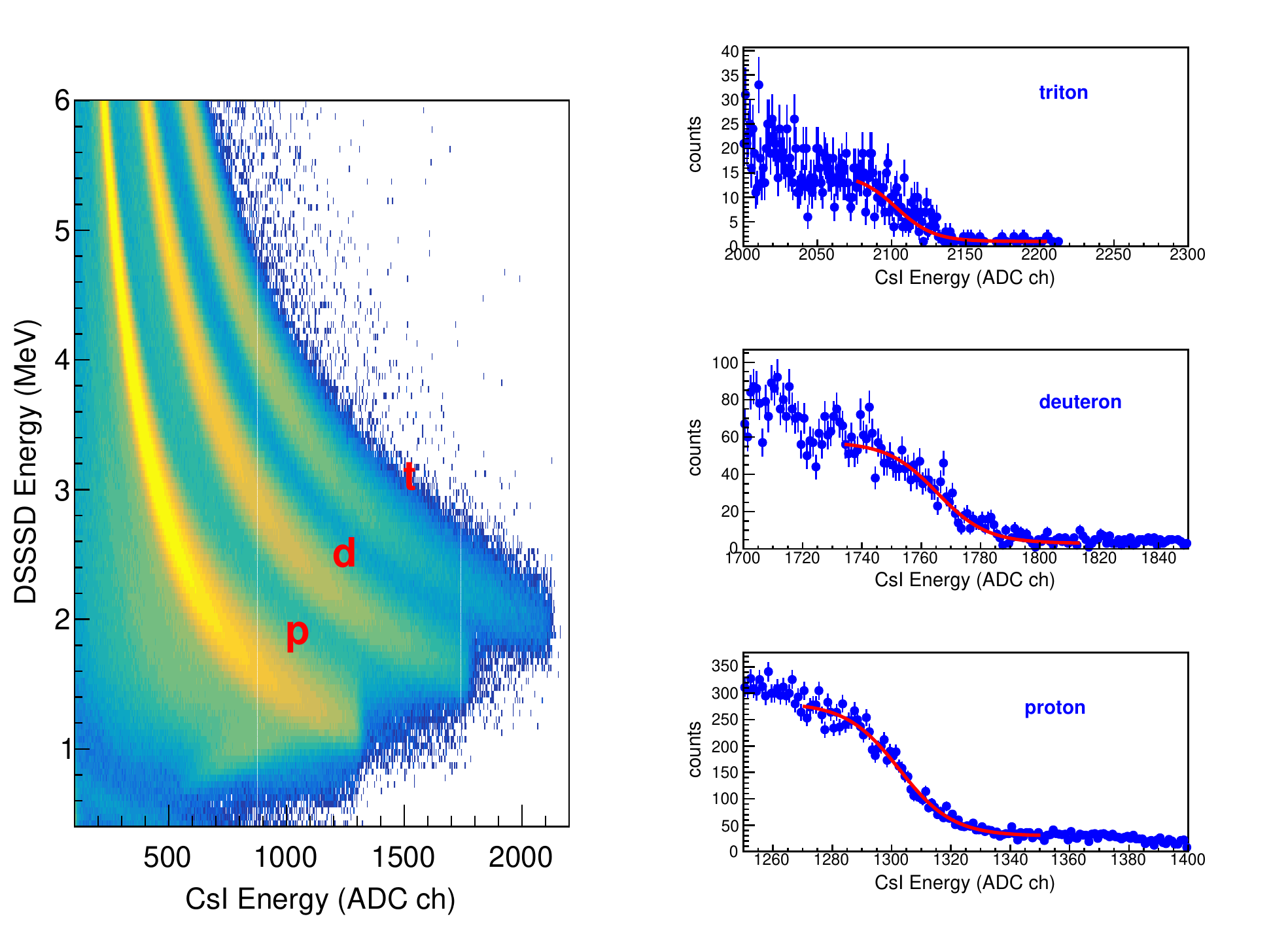}
		\caption{Left panel: proton (p), deuteron (d) and triton (t) DSSSD energy (calibrated in MeV) vs residual energy in a HiRA10 CsI crystal (in ADC channels), zoomed in to the high energy region. The end point of each of the three loci corresponds to the punch-through energy of the corresponding isotope. Right panels: vertical projections of the three loci in the CsI energy axis. The red lines are the result of three individual fits to extract the CsI ADC channel corresponding to the punch-through energy value.}
		\label{figure_05}
	\end{center}
\end{figure} 
The maximum energy that a certain type of incident charged particle can deposit in the crystal is the one for which the particle is completely stopped in the crystal just before punching through.  This is known as the punch-through energy of the particular ion and represents the transition point of the $\Delta$E-E line from the usual shape to the reversed band. The additional reversed band is caused by a particle striking a crystal with a high energy, punching through the entire crystal, and only partially depositing its energy. This is clearly visible in the left panel of Figure~\ref{figure_05} for protons, deuterons and tritons. In the energy-light calibration of a CsI crystal, the punch-throughs represent the end-point or the highest energy point of the calibration curve. A $10\pm0.2$ cm crystal has a punch-through energy of $198.5\pm2.6$ MeV for protons, $263.6\pm3.2$ MeV for deuterons, and $312.4\pm3.3$ MeV for tritons. The light output correspondence of the punch-through of each of the observed hydrogen isotopes is extracted by projecting the corresponding $\Delta$E-E locus on the E axis. The right panels of Figure~\ref{figure_05} show the corresponding distributions for protons, deuterons, and tritons. Figure~\ref{figure_05} is produced with data from a typical crystal located in the intermediate angular range of the HiRA10 cluster. Crystals located at more forward angles show more pronounced drops with higher statistics, while the punch-through of $^3$H is not visible for some of the crystals located at backward angles. The distributions in the right panels are then fitted with the sum of a Fermi function ($f(x)=\frac{a_0}{e^{((x-a_1)/a_2)}+1}$) and a linear term to account for the background. The observed light output corresponding to the particle punch-through is determined by the $a_1$ term derived by the fit procedure, while $a_2$, which represents the sharpness of the distribution, is used to quantify the associated error. We have carefully verified that our chi-square minimization procedure is negligibly affected by the choice of the interval used to perform the fit. This is true if such interval includes a significant flat region before and after the drop due to the punch-through. The $a_2$ parameter is affected by a combination of many factors including the light output resolution of the crystal, possible reaction or scattering of the incident particle in the crystal (resulting in incomplete energy  collection), and the contribution of different beam angles that results in a distribution of different effective-CsI lengths experienced by the incident particle.
\subsection{Direct low-energy beams calibration data}
\begin{figure}[t]
	\begin{center}
		\includegraphics[scale=0.35]{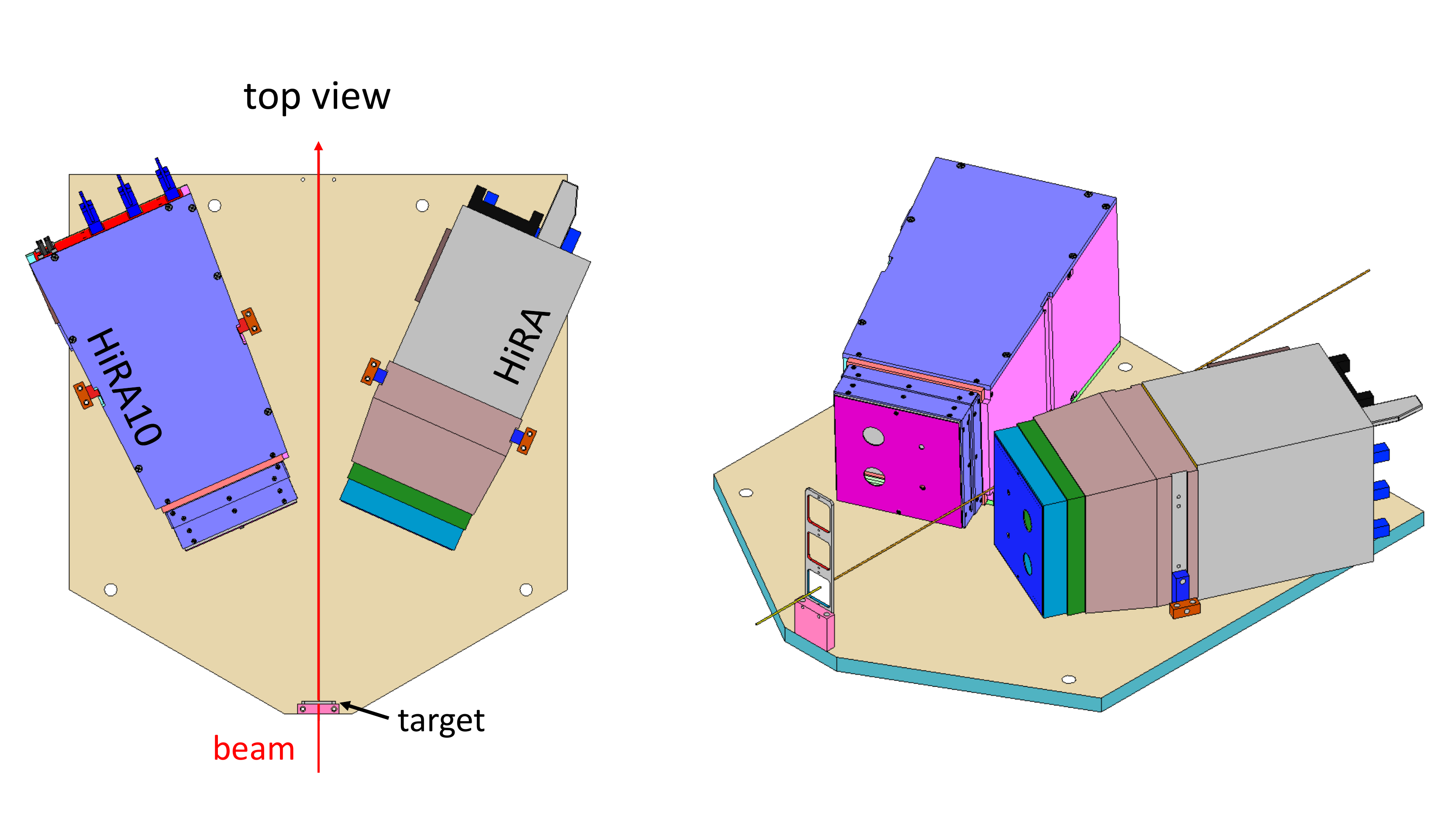}
		\caption{A schematic of the setup used for the WMU experiment. One HiRA10 and one HiRA telescope without DSSSDs are installed at $\pm24$ deg with respect to the beam line (red arrow) as shown on the top view of the vacuum chamber plate (left panel). The collimators placed at the entrance window of each crystal are visible in the right panel, which shows an alternative view of the setup.}
		\label{figure_06}
	\end{center}
\end{figure} 
\begin{figure}[t]
	\begin{center}
		\includegraphics[scale=0.5]{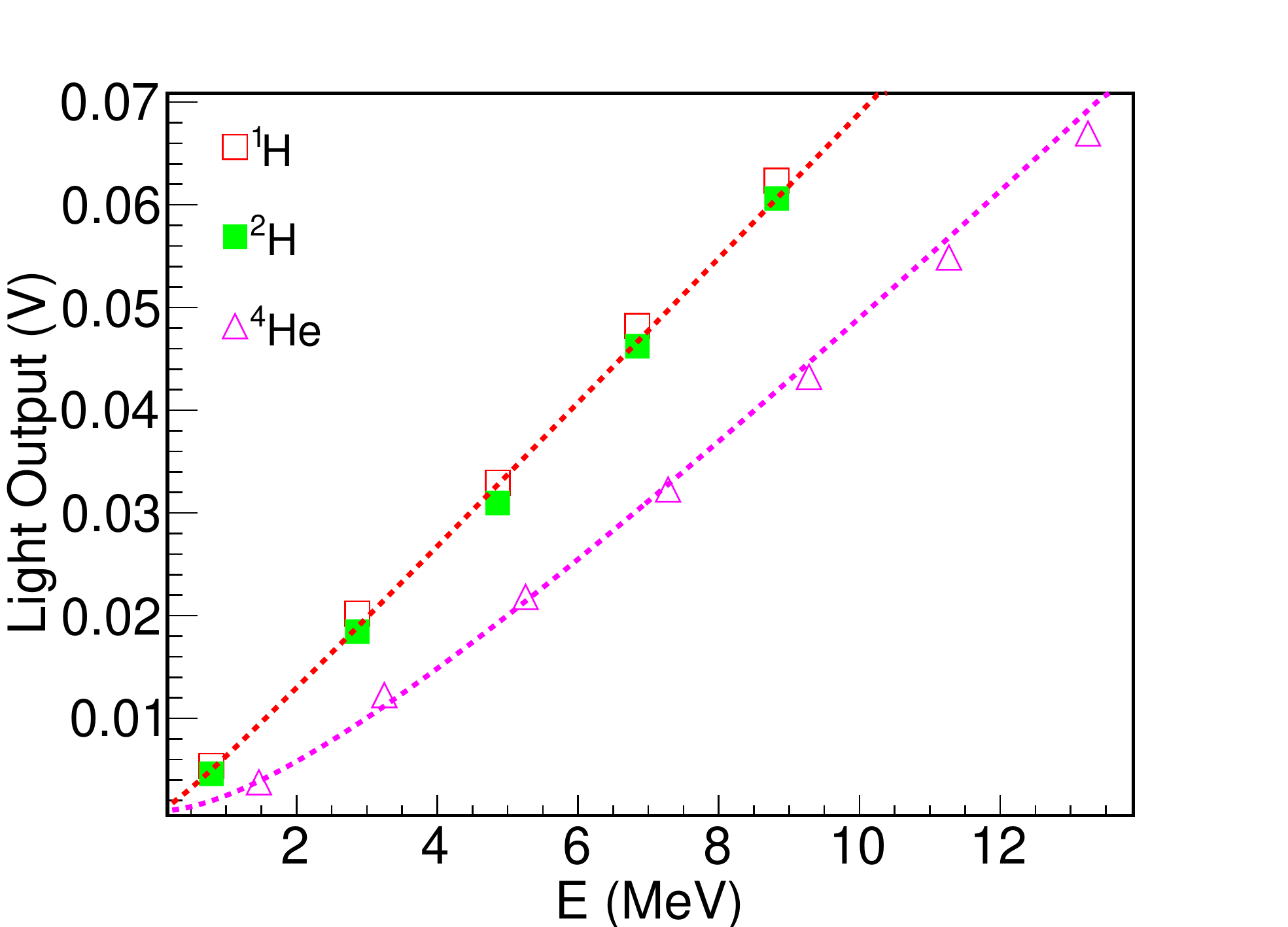}
		\caption{Total CsI light output as a function of the incident energy for protons (red open squares), deuterons (green solid squares) and $^4$He (open triangles) as obtained with our dedicated experiment at WMU, shown for one of the tested HiRA10 crystals. Dashed lines are the results of a simultaneous fit on helium and hydrogen isotopes with the formalism discussed in [Horn92,Laroc94]. Experimental light output points are used to constrain the low energy region of the energy-light calibrations.}
		\label{figure_07}
	\end{center}
\end{figure} 
Finally, to effectively calibrate the lower energy region of hydrogen and helium isotopes, we have performed a dedicated experiment at the Tandem Van de Graaff accelerator facility of Western Michigan University (WMU). For this experiment, one HiRA10 telescope without the DSSSD was placed at $24$ degrees with respect to the beam line in a vacuum chamber. Additionally, a standard HiRA telescope (with $4$ cm CsI crystals) without the DSSSDs, for which the light response has been previously tested at low energy \cite{Man18} with an analogous setup, was placed symmetrically at $-24$ deg with respect to the beam line to serve as the reference for the HiRA10 test. A copper collimator with 4 circular holes centered at the $4$ crystals with diameters of 1/8 in (0.317 cm) and 1/2 in (1.27 cm) for the 2 inner and the 2 outer crystals was placed in front of each HiRA telescope. The hole size accounts for the large differences in the Rutherford scattering at different angles.  This set up, shown schematically in Figure~\ref{figure_06}, allows for simultaneous testing of 8 bare CsI crystals. Proton and deuteron beams at energies of $1$ MeV, $3$ MeV, $5$ MeV, $7$ MeV and $9$ MeV, and $^4$He beams at $2$ MeV, $4$ MeV, $6$ MeV, $8$ MeV, $10$ MeV, $12$ MeV and $14$  MeV  were elastically scattered off a $^{12}$C target with areal density of $107$ $\mu$g/cm$^2$. The center of the light output distribution measured by each of the crystals was extracted by weighting with the Rutherford angular distribution for the corresponding projectile to correct for effects due to the finite angular range covered by the collimators. Energy loss of the incident beam travelling through the target, assuming that the scattering happens at mid target, has been taken into account in calculating the expected energy, as well as the energy loss of outgoing scattered particles through the target and the aluminized Mylar layer at the entrance of the CsI. In both the WMU and the NSCL experiments, the electronic gain of each CsI detection channel was carefully calibrated by using a ramp of $110$ calibrated pulses with a $0.01$ V step size. At the low-end of the electronic dynamic range, a finer step size of $0.002$ V was used to correct with higher precision electronic non-linearities close to the pedestal, which affect the zero-offset. Using the pulser, we can combine data from the WMU experiment with data collected in the NSCL experiment. The result of the WMU experiment for one of the investigated crystals is shown in Figure~\ref{figure_07}. The x-axis reports the calculated scattered energies of the various impinging particles while the y-axis is the calibrated light output in volts. The light response of $^4$He (open triangles) results quenched with respect to the response observed for hydrogen isotopes. The trend is slightly non-linear across the whole energy range explored, as predicted by the light-energy parametrization of the type given in \cite{Horn92,Laroc94}. This trend is quantified by the red (hydrogen) and purple (helium) dashed lines in the Figure, which represent the result of a simultaneous fit of the low energy region of hydrogen and helium obtained by using the standard light-output calibration formula discussed in \cite{Horn92}. For protons and deuterons, (open and full squares), we see a linear trend down to approximately $1$ MeV, indicating a negligible quenching effect. This is probably due to the high concentration of the activator element and, results in a negligible separation between $^1$H and $^2$H. 
\section{Non-linearity effects on the light output} 
\begin{figure}[t]
	\begin{center}
		\includegraphics[scale=0.5]{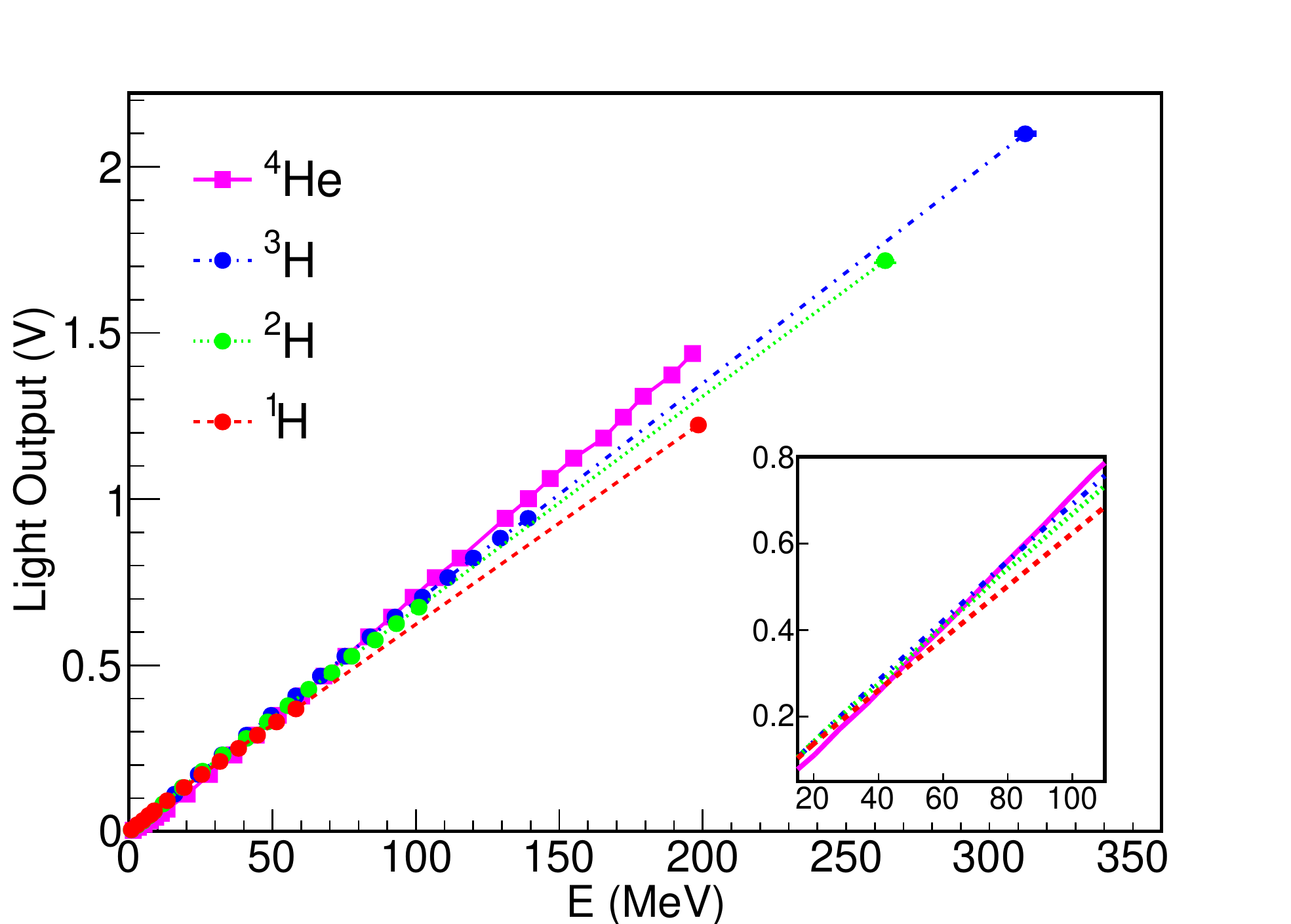}
		\caption{Comparison between hydrogen (red points, $^1$H, green points, $^2$H, blue points, $^3$H) and helium (purple squares, $^4$He only) light output as obtained with the procedure described in Section 2. For clarity reasons, light output points obtained by analyzing proton-recoil scattering data, which are partially in overlap with the points obtained by the energy loss method, are not shown. Dashed (hydrogen isotopes) and solid (helium) lines are used to guide the eye.}
		\label{figure_08}
	\end{center}
\end{figure}

\begin{figure}[t]
	\begin{center}
		\includegraphics[scale=0.5]{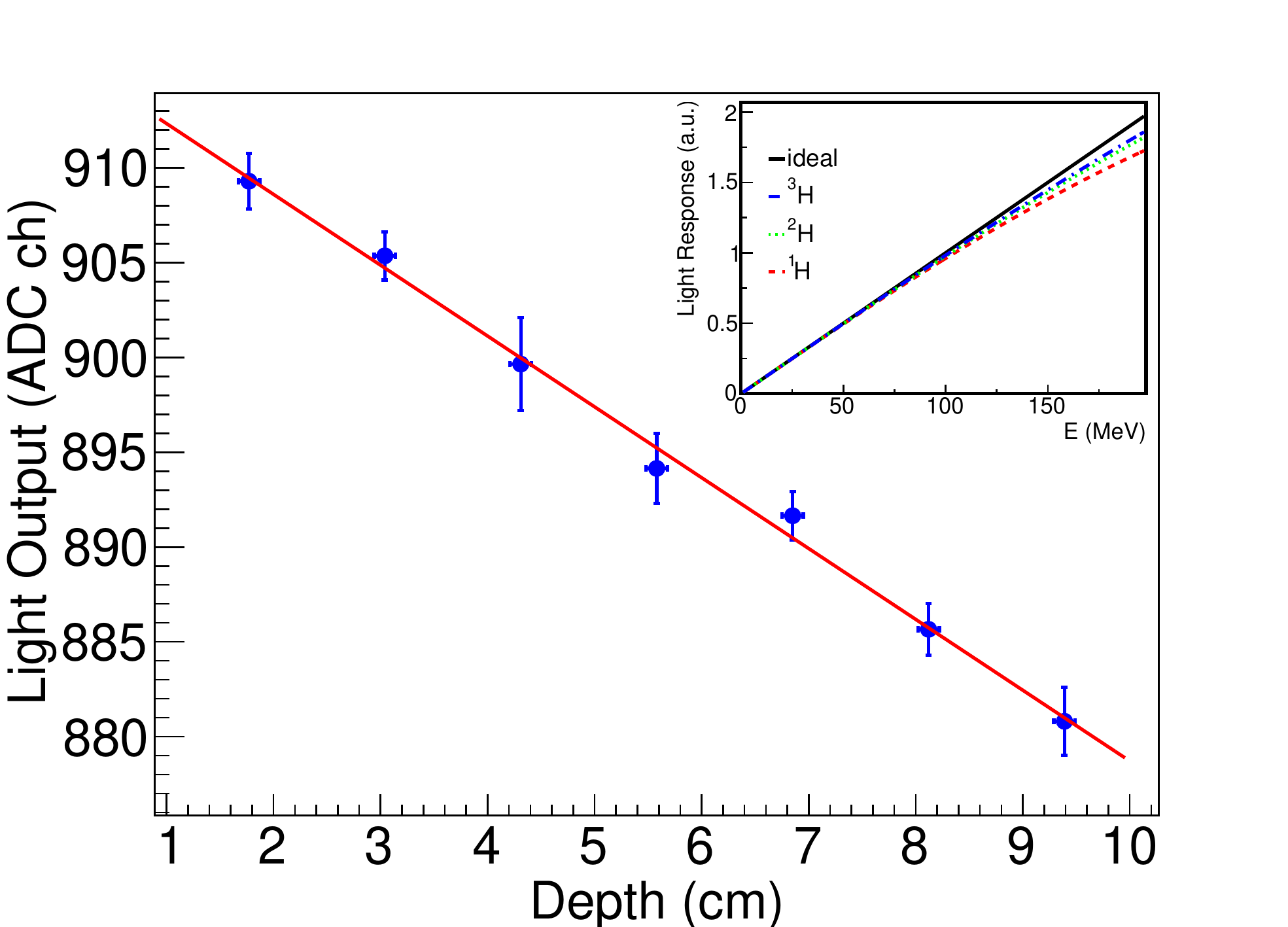}
		\caption{Result of the $^{137}$Cs $\gamma$-ray experiment longitudinally exploring the light output of a HiRA10 crystal. The light output is measured over $7$ points along its length. The result of a linear fit is also shown. The top right panel shows how an ideal light response of equation~\ref{eq2} is modified because of the considerations of equation~\ref{eq4}. Calculations are shown for $^1$H (red dashed line), $^2$H (green dotted line) and $^3$H (blue dash dotted line).}
		\label{figure_09}
	\end{center}
\end{figure}

We briefly summarize the features observed in Figure~\ref{figure_07}, which serve as the starting point for the following discussion: (1) the measured light output for $Z=2$ isotopes is consistently lower with respect to the one measured for $Z=1$ isotopes across the entire energy range; (2) the helium light output is slightly non-linear; (3) the hydrogen-isotope light output does not exhibit significant non-linearities and a negligible separation between $Z=1$ isotopes is observed, with a slightly larger light response for $^1$H with respect to $^2$H. These facts are clearly in unquestionable agreement with the predictions of conventional energy-light parametrizations for light ions \cite{Horn92,Laroc94}. To better understand the meaning of this statement, let us consider the following equation derived by the Birks formalism for inorganic scintillators:
\begin{equation}
  dL/dE=  \frac{S}{1+KB|dE/dx|}
  \label{eq1}
\end{equation}
Here $S$ is the scintillation efficiency, which corresponds to the amount of light produced per unit of energy released in the crystal in the absence of quenching, $dE/dx$ is the particle stopping power and $KB$ is the so-called quenching factor. This equation gives the differential scintillation efficiency per unit of energy. For sufficiently high energies, it results in an almost constant value because of the small stopping power, while the light efficiency is reduced at lower energies, resulting in a non-linear integrated light output with respect to the deposited energy. For sufficiently small quenching factors, as in the case of highly-doped crystals, $KB|dE/dx|$ does not play a significant role in $Z=1$ isotopes, and the corresponding light response  is nearly linear down to very low energies as observed in Figure~\ref{figure_07}. Because of the larger stopping power of $Z=2$ isotopes with respect to $Z=1$ at all energies, $\left(dE/dx\right)_{(Z=2)}>\left(dE/dx\right)_{(Z=1)}$, the quenching term plays a role in the $Z>1$ light response. In particular, one would expect $\left(dL/dE\right)_{(Z=2)}<\left(dL/dE\right)_{(Z=1)}$ and therefore $\int{\left(dL/dE^\prime\right)_{(Z=2)}dE^\prime}=L(E)_{(Z=2)}<\int{(dL/dE^\prime)_{(Z=1)}dE^\prime}=L(E)_{(Z=1)}$. In other words, if $S$ and $KB$ are constants, for an equivalent energy, the total light response of an inorganic crystal to a helium isotope $L(E)_{(Z=2)}$ is lower than the total light response to a hydrogen isotope $L(E)_{(Z=1)}$ as demonstated in Figure~\ref{figure_07}, representing our observed light output at low energy. However, as shown in Figure~\ref{figure_08}, the light output for hydrogen isotopes constrained with the present analysis shows non-linearities that result in $L(E)_{(Z=1)}<L(E)_{(Z=2)}$ at energies higher than $\approx50$ MeV, a trend more similar to the one described in Refs.~\cite{Meijer87,Twen90,Aiello96}. In Figure~\ref{figure_08}, the energy-light constraints obtained with the energy-loss method and the WMU calibrations are shown for hydrogen isotopes and $^4$He for one crystal by using points connected by dashed (hydrogen)  and solid (helium) lines to guide the eye. In the figure, proton-recoil kinematic points were not included for the sake of clarity, as they partially overlap with the other $^1$H data. For the same reason, due to the small separation between $^3$He, $^4$He and $^6$He, only $^4$He is shown as the representative of Z=2 isotopes. Another interesting fact, visible in the inset of Figure~\ref{figure_08} where the symbols indicating the experimental points are removed for the sake of clarity, is the systematic inversion of $^1$H, $^2$H and $^3$H, having $L(E)_p\approx L(E)_d\approx L(E)_t$ at low energy and $L(E)_p<L(E)_d<L(E)_t$ at energy higher than about $20$ MeV. Therefore a conventional light output calibration is not applicable to the case of light isotopes, and additional contributing factors affecting the response and light collection of our crystals have to be carefully taken into account to understand and correctly calibrate the observed trends.
To quantitatively understand experimental results described above we develop a simple model which will show that these effects can be mainly attributed to the spatial non-uniformity of the crystal related to the Thallium doping and the light collection efficiency. Let us assume that the light response of the crystal is not affected by quenching ($KB=0$) and that the scintillation efficiency is a constant in each point of the crystal. In the presence of an ideal light collection, i.e. all the produced optical photons are collected at the photodiode without absorptions or leaks, the measured light output is given by the simplified formula
\begin{equation}
  L(E)=\int_{E}^{0}{S\frac{dE}{dx^\prime}dx^\prime}=S\int_{E}^{0}{\frac{dE}{dx^\prime}dx^\prime}=SE
  \label{eq2}
\end{equation}
This formula is often used in the literature to calibrate the light response of hydrogen isotopes assuming negligible quenching effects \cite{Wag01}. However, the quantity S is usually not a constant since it contains a spatial dependence of the type $S(x,y,z)$=$\epsilon(x,y,z)\rho(x,y,z)$, where $\epsilon(x,y,z)$ is the light collection efficiency, i.e. the probability that a photon emitted at $(x,y,z)$ is effectively detected by the photodiode, ($\epsilon(x,y,z)$ is typically of the order of few tens per cent) and $\rho(x,y,z)$ is the light production efficiency, which is related to the density of the activator element at the position $(x,y,z)$ of the crystal bulk. Disentangling $\epsilon(x,y,z)$ and $\rho(x,y,z)$ is extremely complicated, since the first has to be carefully calculated and is sensitive to a number of factors such as the refraction index of various materials, surfaces, optical couplings, etc. However, the investigation of their product can be done experimentally. To this end, we have longitudinally scanned our crystals with a collimated $^{137}$Cs gamma source. A thick lead collimator was used to provide a focused source of mono-energetic gamma rays ($E_\gamma=662$ keV) at various points along the length of the crystal. The source was placed on one side of the crystal perpendicularly to its axis. Measurements of the $\gamma$ full energy peak were performed at 7 different locations along the length of the crystal. Results of these measurements are shown in Figure~\ref{figure_09}, where the measured light output is shown as a function of the source position along the crystal. The experimental data exhibit a linearly decreasing trend, as shown by the red fit line. If we introduce the assumption that $S(x,y,z)$ is a function uniquely of the longitudinal position within the crystal, $S(x,y,z)=S(z)$, one can express the spatial dependence of the scintillation efficiency with the following simplified equation
\begin{equation}
  S(z)=a(1-bz)
  \label{eq3}
\end{equation}
where $a$ is a gain factor and $b$ is the light-output gradient revealed by our gamma-source investigation. It is important to stress that the hypothesis of uniformity of the crystal along surfaces orthogonal to the axis is quantitatively supported by the observed uniformity of the crystals at the entrance surface and at the base (see Section 2). Under this simplification, and by considering, for simplicity, a particle traveling longitudinally, equation~\ref{eq2} can be rewritten into a more general form
\begin{equation}
  L(E)=\int_E^0{S(z)\frac{dE}{dz}dz}=a\int_E^0{(1-bz)\frac{dE}{dz}dz}=aE-b\int_E^0{z\frac{dE}{dz}dz}
  \label{eq4}
\end{equation}
which describes the energy-light relation for a light particle ($KB=0$) striking the crystal orthogonally to the entrance surface in the presence of a position-dependent light output. Apart from the linear term of the equation~\ref{eq2}, an additional term is present which makes the expected light output non-linear. The corresponding differential efficiency dL/dE decreases for increasing energies, as qualitatively observed in our data. For a more quantitative analysis, equation~\ref{eq4} can be numerically integrated. This results in the approximated analytical expression
\begin{equation}
  L(E)\approx a^\prime E^\gamma
  \label{eq5}
\end{equation}
where $a^\prime$ is a gain factor and $\gamma$ depends on the nature of the incident ion, being $\gamma_p<\gamma_d<\gamma_t<1$. The top right panel of Figure~\ref{figure_09} shows the results of a similar calculation for protons, deuterons and tritons, compared to the original light-output of equation 2. A saturating behavior analogous to that observed in the experimental data of Figure~\ref{figure_08} is clearly produced, justifying our assumption.

\begin{figure}[t]
	\begin{center}
		\includegraphics[scale=0.5]{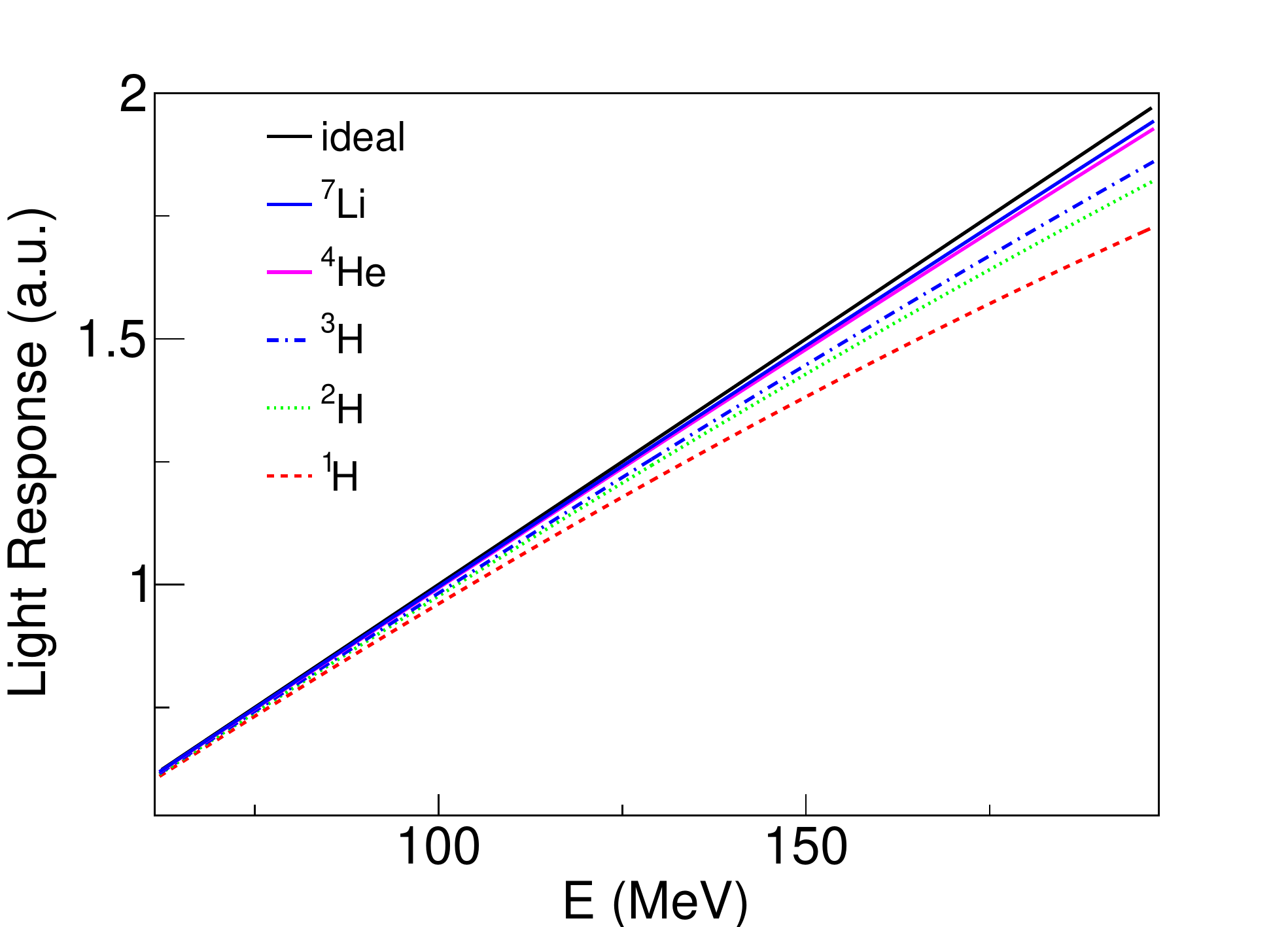}
		\caption{Comparison of a simulated ideal light response of a CsI, obtained by using equation~\ref{eq2}, with the result of equation~\ref{eq4} for several incident ions: $^1$H (dashed red line), $^2$H (dotted green line), $^3$H (dash dotted blue line), $^4$He (purple line), $^7$Li (blue line). Under our simplified hypothesis,  a saturating behavior is observed to play a significant role in this energy range only in the case of hydrogen isotopes. Slight deviations from the linearity are also observed for helium and lithium isotopes.}
		\label{figure_10}
	\end{center}
\end{figure}

The last crucial point of the present considerations relates to the extent to which the described effects affect the light output of ions of different nature. As pointed out by equation~\ref{eq5}, a saturating trend like the one observed in Figure~\ref{figure_08} is compatible, within the limits of our simplified hypothesis, with our argument. Anyway, since the saturation parameter $\gamma$ is affected by the spatial properties of the crystal, it is intimately related to the range of a particle in the crystal. To better clarify this statement, let us consider the following points. A simple consideration of the dependence of a particle’s stopping power on its energy indicates that most of the light produced by a particle in its path through the crystal is produced close to the end of the track as a result of the rapid increase of the stopping power of ions at lower energies. This makes these results strongly dependent on the range of the radiation through the crystal. Figure~\ref{figure_10} better indicates this point by comparing the result of our simulation, obtained with a realistic $b\approx0.2 cm^{-1}$ factor, for hydrogen, helium and lithium isotopes and the ideal curve obtained by equation~\ref{eq2}. The plot shows an extremely small saturation effect for $Z\geq2$ with respect to the one obtained for hydrogen, confirming the reduced sensitivity of heavier ions to spatial non-uniformities of the crystal. To be more quantitative, a proton at an energy of $100$ MeV will produce light in a CsI crystal along a $3.1$ cm-path, while a $^4$He will travel only for $2.4$ mm before it is completely stopped in the crystal. The two numbers differ by more than a factor of ten. Because of such a macroscopic difference, a crystal will appear spatially uniform to a $^4$He in this energy range while it will show non-uniformities to an incident proton of the same energy. Numerically, we obtain $\gamma\approx0.95$ for protons. However, the $b$ gradient needed to reproduce the trend of the experimental data is larger than the one measured by the gamma-source experiment. This could be caused by the different sensitivity of the light efficiency to a gamma and a charged-particle probe, as observed in Ref.~\cite{Gong88}, and by the different distribution of optical photons produced for a particle longitudinally or perpendicularly penetrating the crystal. Additionally, the possible presence of ballistic deficit \cite{gal95} in the processed signals, caused by the relative short shaping time used in the experiment ($\tau\approx3\mu s$), needed for high-rate applications, might contribute to enhance the hydrogen to helium non-linearity effect observed here. Future investigations of the light output in different regions of the crystal with charged-particle probes are clearly needed. As will be shown in the next section, the non-linearity effects here explored will require a correction term in the light output calibration of hydrogen isotopes. Such a term is found not to play a significant role for heavier elements in the energy range explored here.
\section{Energy calibration}
\begin{figure}[t]
	\begin{center}
		\includegraphics[scale=0.5]{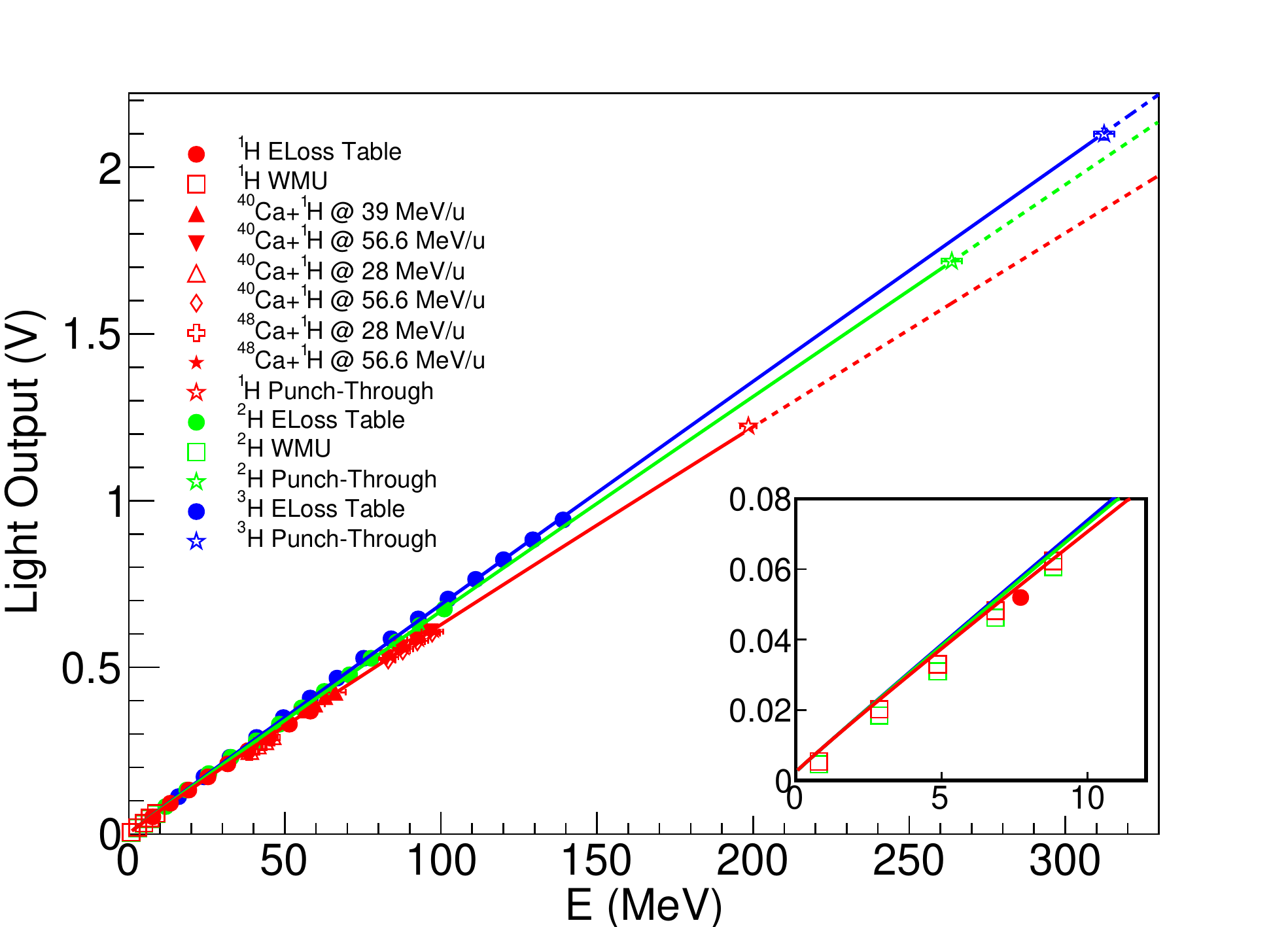}
		\caption{Hydrogen $^1$H (red points), $^2$H (green points), and  $^3$H (blue points) light output calibrations for one of the crystals as constrained by the present work. Different symbols indicate different data sets described in Section 2 of the paper: energy-range method (solid circles), WMU experiment (open squares), proton-recoil kinematics (solid and open triangles, open diamonds, open crosses, solid stars), punch-through (open stars). Fitting lines are obtained by a simultaneous fit with the light output parametrization of equation~\ref{eq6}. Dashed lines indicate the extrapolated light output in the energy region outside of the dynamic range of the crystal for each isotope. The bottom right inset shows a zoom of the low energy region constrained by the WMU experiment.}
		\label{figure_11}
	\end{center}
\end{figure}
\begin{figure}[t]
	\begin{center}
		\includegraphics[scale=0.5]{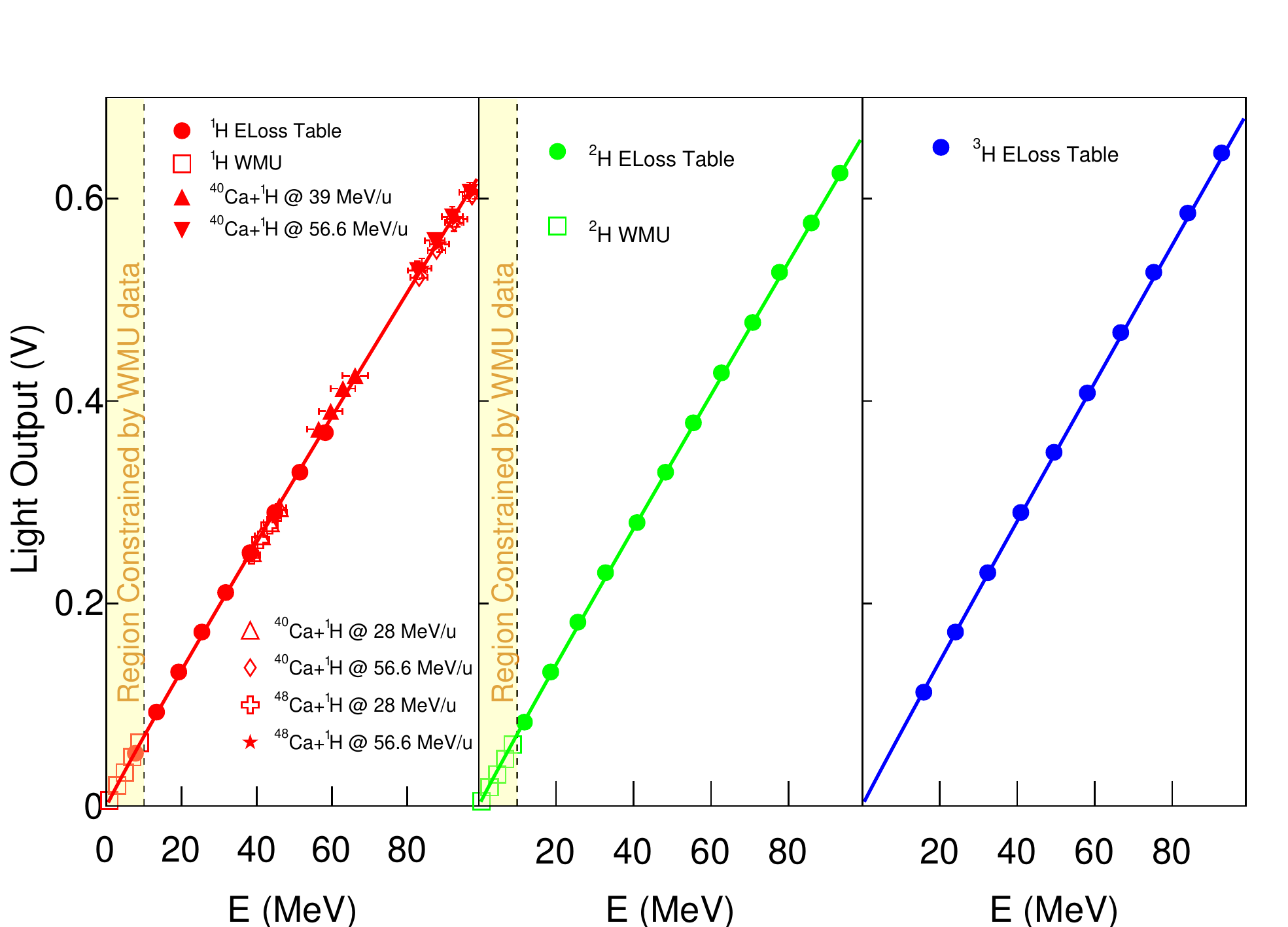}
		\caption{Low energy zoom of Figure~\ref{figure_10} shown separately for $^1$H  (left panel, red color), $^2$H (central panel, green color) and $^3$H (right panel, blue color). The region constrained by the WMU experiment is shown with a shadowed region in both left and central panels. Symbols are the same as in Figure~\ref{figure_10}.}
		\label{figure_12}
	\end{center}
\end{figure}

 
The light output calibration has been obtained by using all the data sets described in Section 2 and by taking into account the considerations of Section 3. We used two separate energy-light calibration formulas, one for hydrogen (that is significantly affected by CsI crystal non-linearity in the relevant energy range) and one for heavier ions (for which the effects described in Section 3 do not play a considerable role). Fit parameters are constrained by using energy-light data of $^1$H, $^2$H, $^3$H, $^3$He, $^4$He, $^6$He, $^6$Li, $^7$Li, $^8$Li, $^7$Be, and $^9$Be, produced with high statistics in the angular domain of the present investigation, while the calibration for other isotopes not included in the fit can be obtained as a simple extrapolation to other Z and A values of the used fitting formula. 
Figures \ref{figure_11} and \ref{figure_12} show the energy-light calibration produced for hydrogen isotopes. To account for the observed non-linearity effects mainly caused by the crystal non-uniformity, we have used a fitting formula derived from equation~\ref{eq4}. The dependence on the ion mass is described by the following empirical parametrization
\begin{equation}
  L(E,Z=1,A)=a_0 E^{\left(\frac{a_1+A}{a_2+A}\right)}
  \label{eq6}
\end{equation}
In the equation, $a_0$ is a gain factor, $A$ represents the $Z=1$ isotope mass number and $a_1$, $a_2$ are empirical non-linearity parameters that satisfy the relation $a_2>a_1$. The best fit is obtained by simultaneously fitting all hydrogen isotopes. Results of the fit are shown for one crystal in Figure~\ref{figure_11} with the three solid lines ($^1$H red line, $^2$H green line, $^3$H blue line). Three dashed lines indicate the extrapolation of the calibration lines in the energy regions outside the dynamic range of the crystal. As clearly visible, we are able to consistently describe the light output of hydrogen isotopes with a unique set of parameters in the whole energy range. The calibration is compatible, within error, with the particle punch-through energies, a key constraint since they define the end point of the light output calibration and lie in a high-energy region more strongly affected by crystal non-linearities. Figure~\ref{figure_12} shows a zoom to the low energy region of Figure~\ref{figure_11} separately for $^1$H (left panel, red color), $^2$H (central panel, green color) and $^3$H (right panel, blue color). The proton energy-light calibration line is also in satisfactory agreement with the constraints obtained from the analysis of proton elastic-scattering recoil data at various energies, as shown in the left panel of Figure~\ref{figure_12}. Inverted full triangles and open diamonds represent kinematic points from $^{40}\textrm{Ca}+^1\textrm{H}$ proton-recoil data measured at $56.6$ MeV/u over two different periods of the experimental campaign to test the stability of the energy-light calibration. The data from these two periods are compatible with each other, suggesting negligible gain change throughout the experiment. In the angular range spanned by the crystal, three groups of kinematic points measured at $28$ MeV/u, $39$ MeV/u and $56.6$ MeV/u contribute in the fit (as shown by Figure~\ref{figure_04} of Section 2). Their different error bars reflect the quality of the beam transport to the experimental target achieved with each individual beam tune. Such uncertainties were carefully investigated and taken into account as discussed in Section 2. Finally, the low energy region (shaded region in Figure~\ref{figure_12}) is well-constrained by the results of the WMU experiment. The WMU data show good agreement with data obtained in the NSCL experiment and a vanishing zero-offset, as visible in the low energy zoom shown in the inset of Figure~\ref{figure_11}. It is important to stress that the linear hypothesis made in previous hydrogen light output calibrations performed with $4$-$cm$ HiRA crystals \cite{Wag01} (where the smaller dynamic range and the limited calibration points did not allow for a detailed study of non-linearity effects) resulted in an incorrect determination of the zero offsets.
\begin{figure}[t]
	\begin{center}
		\includegraphics[scale=0.5]{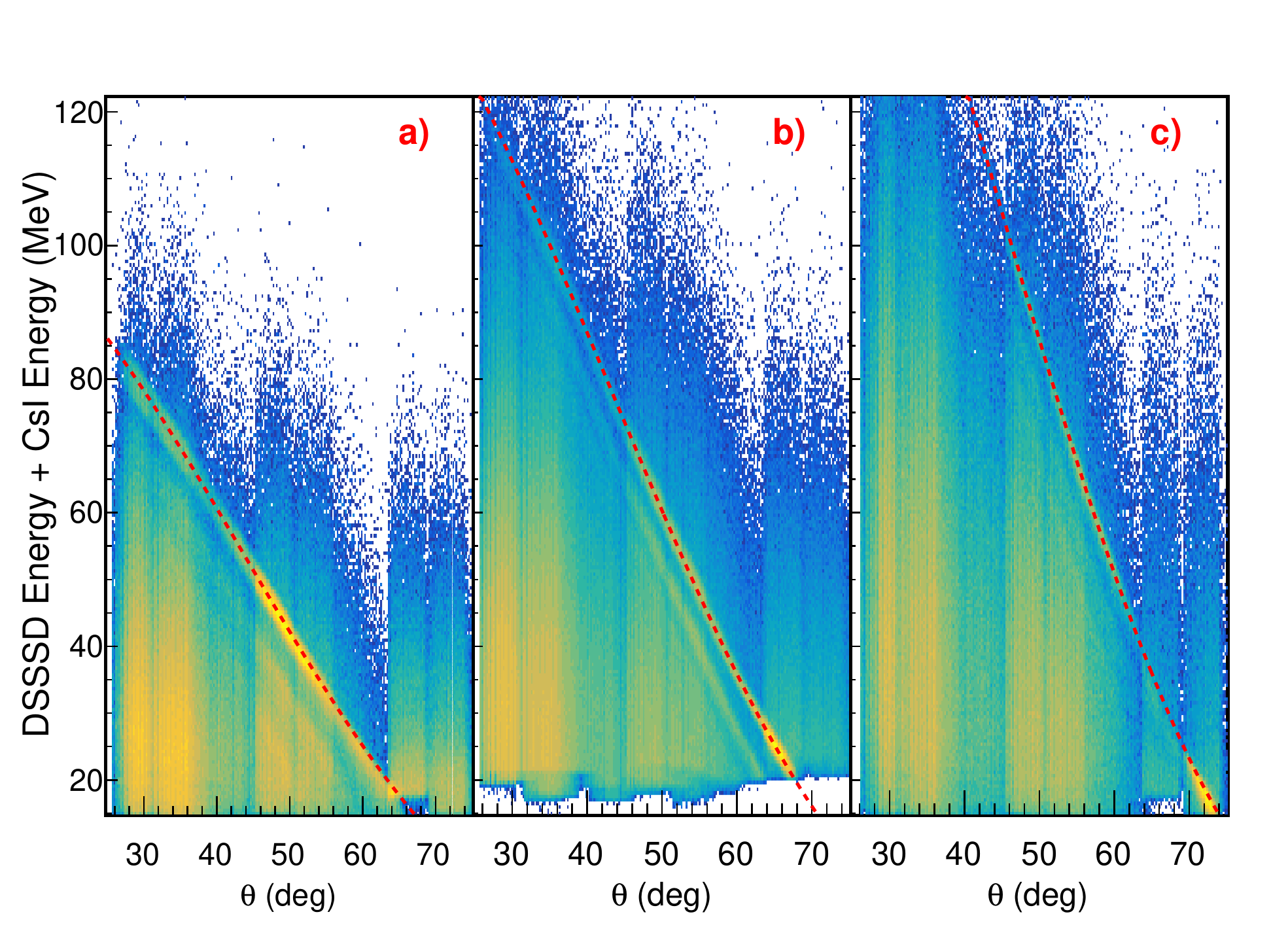}
		\caption{Calibrated proton recoil kinematic lines observed with the full cluster of $48$ HiRA10 crystals in the NSCL experiments, covering a region of polar angles from $25$ to $75$ degrees in the laboratory frame, for a) $^{48}\textrm{Ca} + ^{1}\textrm{H}$ at $28$ MeV/u, b) $^{40}\textrm{Ca} + ^{1}\textrm{H}$ at 39 MeV/u and c) $^{48}\textrm{Ca} + ^{1}\textrm{H}$ at 56.6 MeV/u collisions. The y-axis represents the total kinetic energy reconstructed as the sum of the energy released in the DSSSD and the residual CsI energy. A dashed line is the result of a kinematic calculation that takes into account the energy losses of the protons in the target and the foils placed in front of the DSSSD entrance window.}
		\label{figure_13}
	\end{center}
\end{figure} 
To demonstrate the validity of our approach for all crystals in the array, Figure~\ref{figure_13} shows the same proton-recoil kinematic lines as described by Figure~\ref{figure_04} but with data combined from the full cluster of $48$ HiRA10 crystals, covering the angular range shown on Figure~\ref{figure_01} (right panel). This is a particularly useful check to test the validity of the produced energy calibrations. In the figure, the Y-axis is the calibrated kinetic energy of protons at the entrance of the telescopes, calculated as the sum of the calibrated energy loss in the DSSSD and the residual energy in the CsI crystal. A marked ridge, ranging from polar angles of $25$ to $75$ degrees is clearly visible, indicating the overall internal consistency of our hydrogen calibrations for all $48$ crystals. The background is caused by nuclear reactions and  inelastically scattered protons. We have performed a theoretical calculation of the energy of elastically scattered protons for the three collision systems shown in the figure: a) $^{48}\textrm{Ca}+^1\textrm{H}$ at $28$ MeV/u, b) $^{40}\textrm{Ca}+^1\textrm{H}$ at $39$ MeV/u and c) $^{48}\textrm{Ca}+^1\textrm{H}$ at $56.6$ MeV/u. The calculation includes also the energy loss of the outgoing protons through the target and the absorber foils placed in front of the DSSSD entrance window of each telescope. The red lines show an overall good agreement with the experimental lines that correspond to the elastic scattering events for all the collision systems shown here. The extra lower energy line clearly visible in panel b) is produced by inelastic scattering events that leave the scattered $^{40}$Ca in one of its first close-lying excited states ($3.35$ MeV $0^+$, $3.74$ MeV $3^-$, $3.90$ MeV $2^+$). For the higher energy system (panel c), we cannot observe a marked kinematic line at forward angles because of the reduced cross section at this energy. The slight energy shift observed between the theoretical and experimental lines at large detection angles for the lower energy (panel a) is mainly because of the higher angular uncertainty of the HiRA10 pixels due to the larger beam spot on the target achieved for the $28$ MeV/u beam during the NSCL experiment. This affects the effective angles at which particles are detected, which has a more pronounced effect for backward detection angles due to the inverse kinematics.

\begin{figure}[t]
	\begin{center}
		\includegraphics[scale=0.5]{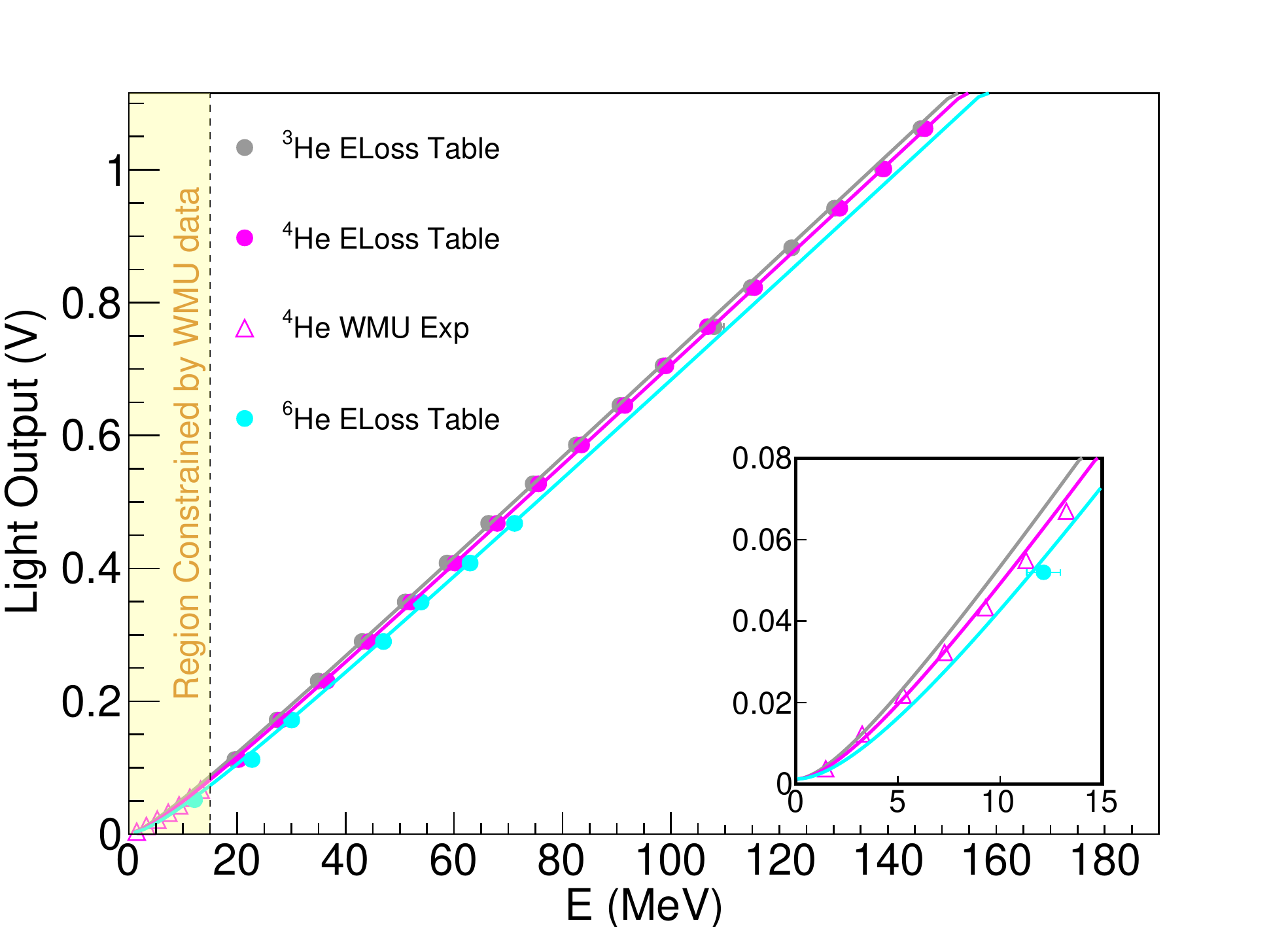}
		\caption{Light output calibration for helium isotopes $^3$He (gray points), $^4$He (purple points), $^6$He (light azure points) for one of the analyzed crystals. Solid circles are obtained by the energy-loss method while open triangles represent data from the WMU experiment ($^4$He), covering the shadowed region in the figure. An inset shows a zoom of the low energy region constrained by WMU data. The dynamic range of $^6$He is reduced because of the collected statistics in the angular region covered by the crystal.}
		\label{figure_14}
	\end{center}
\end{figure}
\begin{figure}[t]
	\begin{center}
		\includegraphics[scale=0.5]{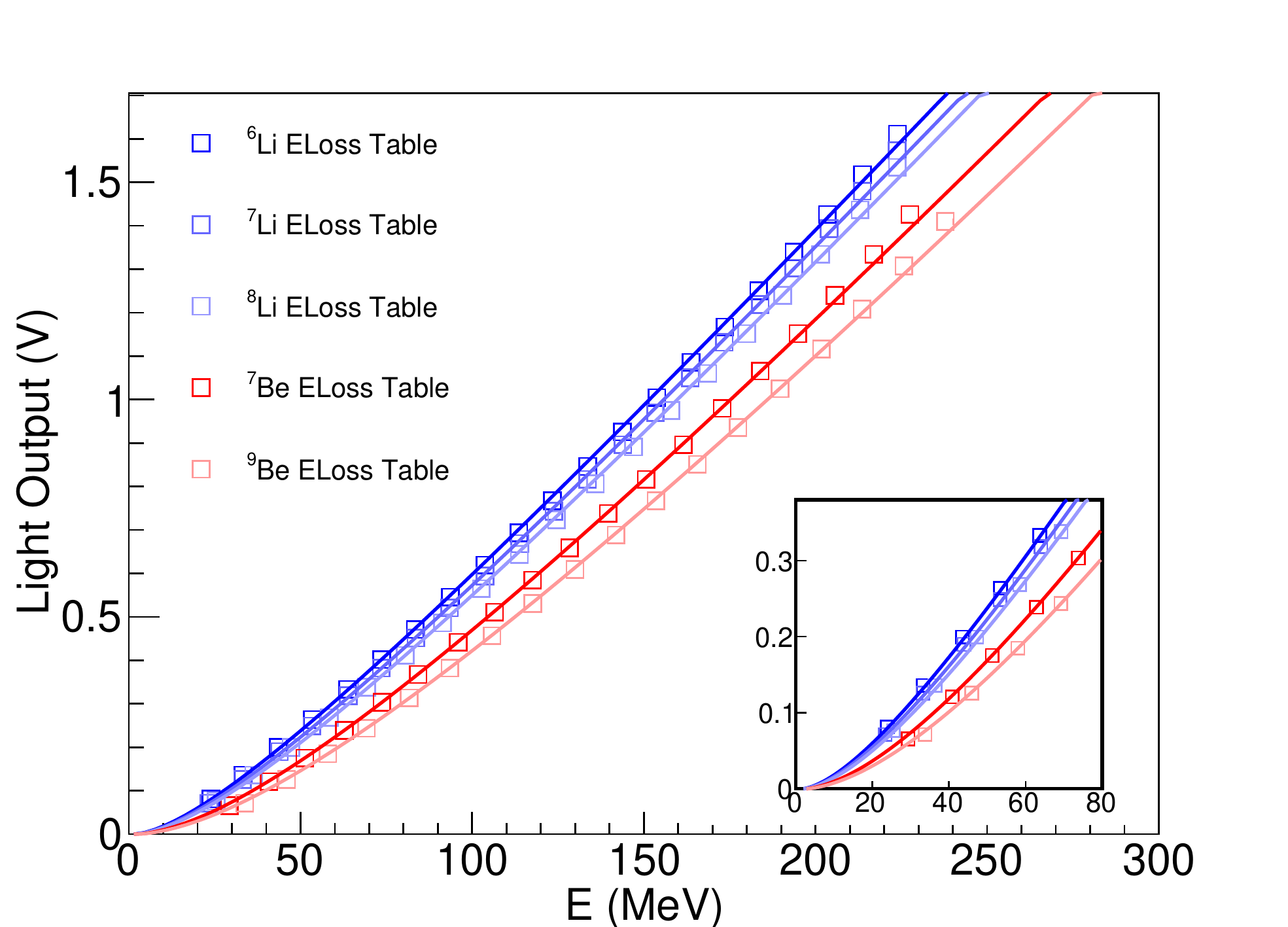}
		\caption{Light output calibration obtained for heavier isotopes: $^{6,7,8}$Li (different tones of blue) and $^{7,9}$Be (different tones of red). All the calibration points considered for such isotopes are obtained by the energy loss method. An inset, shown in the bottom right corner, represents a zoom of the lower energy region.}
		\label{figure_15}
	\end{center}
\end{figure}

 
The energy-light calibration of heavier isotopes is obtained by using the standard light-output parametrization \cite{Horn92} obtained by an analytical integration of equation~\ref{eq1} under the approximation $dE/dx \propto -AZ^2/E$:
\begin{equation}
  L(E,Z>1,A)=a_0\left(E-a_1AZ^2log\left(\frac{E+a_1 AZ^2}{a_1 AZ^2}\right)\right)
  \label{eq7}
\end{equation}
where $a_0$ and $a_1$ are parameters obtained from a simultaneous fit of data. The expression presents a linear part dominating at high energies which is characterized by a slope given by $a_0$ and is related to the light efficiency of the crystal. There is also a non-linear quenching term dominating the lower energy region and affected by the fitting term $a_1$. The additional $AZ^2$ dependence accounts for the different quenching because of the different mass and charge of the impinging ions. Figures \ref{figure_13} and \ref{figure_14} show the result of the fit to helium and heavier isotopes, respectively. Additionally, the inset in Figure~\ref{figure_14} shows a low energy zoom of the region constrained by WMU data, also seen in the full-scale figure. A clear isotopic separation in the light output is visible both in the case of helium isotopes and in the case of heavier isotopes and is found to be in good agreement with the light output parametrization of equation~\ref{eq7}. The limited energy range of $^6$He reflects the low statistics of production of this isotope in our  experiments. The non-linear trend of equation~\ref{eq7} at low energy reproduces well the results of the WMU experiment with small uncertainties, mainly from a small gain mismatch due to the electronics calibration in volts in the two different experiments. At higher energies, the light output observed for helium isotopes does not show any significant non-linearity, as expected from the considerations of Section 3 and attributed to their shorter range in the CsI crystal in the relevant energy domain. A good fit of data is obtained up to energies of around $250$ MeV. Higher energy calibration points are dominated by uncertainties arising from the energy-loss method and the lower statistics. The constraints used to calibrate the light output of heavier isotopes are more limited, since the angular range covered by the crystals during the experiment was optimized for the detection of light ions. Among the crystals in the cluster, the one used to produce the figures described in the paper constitutes a typical example of a crystal located in the intermediate angular range of the cluster, as described in the right panel of Figure~\ref{figure_01}. In this case, shown in Figure~\ref{figure_15}, calibration points are uniquely extracted by means of the energy-loss method, using the multi-parametric fitting procedure described in Section 2 and adopted for $Z>2$ isotopes, with the result of larger systematic errors in the high-end of the dynamic range explored in the calibration. For this reason, calibration points have been limited to energies of around $250$ MeV. Also, in this case, the quenching light response is found to produce a good fit to the data. The explicit mass and charge dependence of equation~\ref{eq7} allows for application of the same calibration to other isotopes not included in the fit because of their very limited statistics, but well identified with the $\Delta$E-E technique.

\begin{figure}[t]
	\begin{center}
		\includegraphics[scale=0.5]{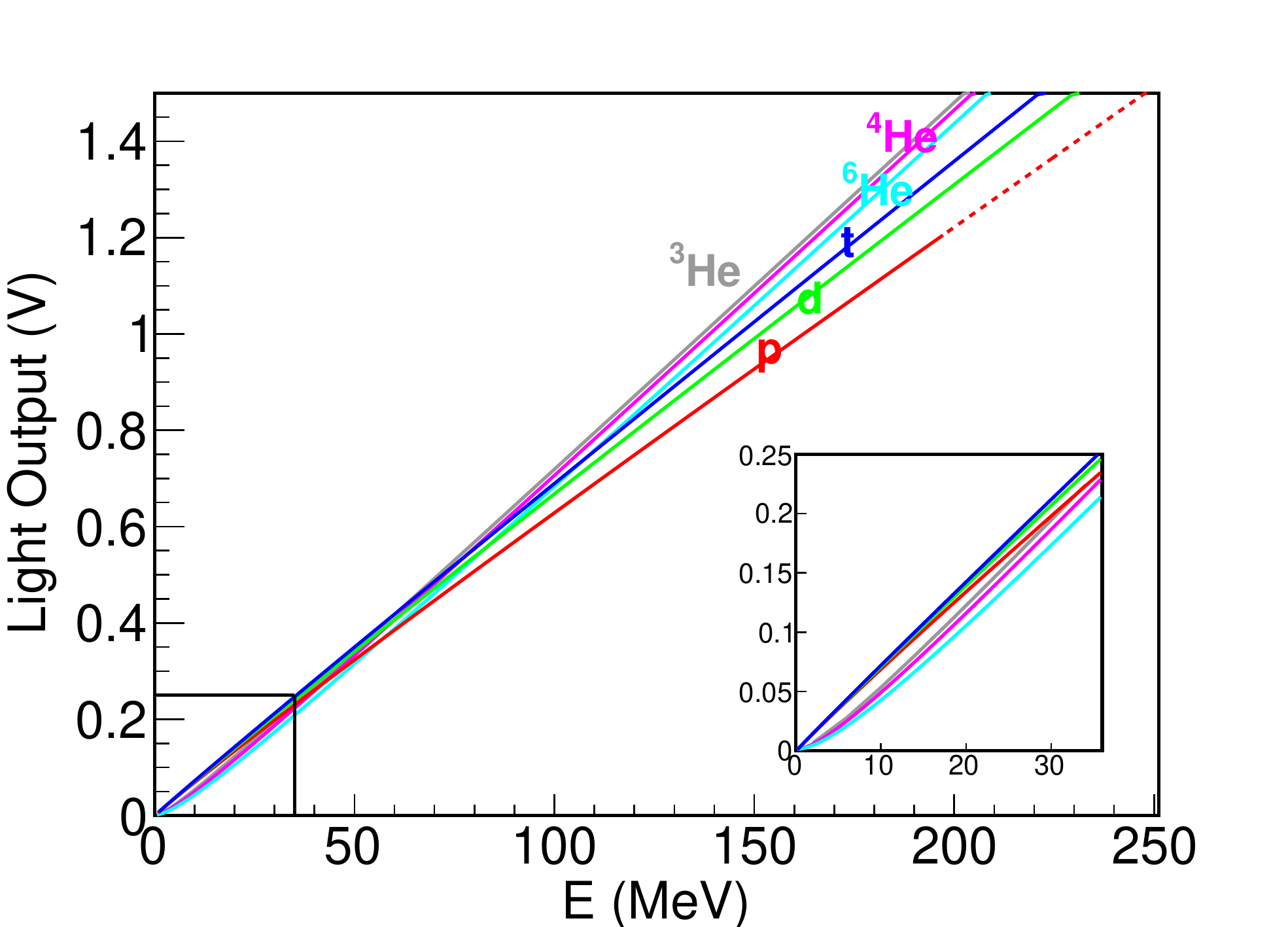}
		\caption{Energy-light calibration curves as obtained in the present paper by using, respectively, the mass dependent calibration of equation~\ref{eq6} and the charge and mass dependent calibration of equation~\ref{eq7}. The dashed line is the extrapolation of the calibration line outside of the range of the crystal. The bottom right inset is a zoom of the region within the black box shown on the figure to emphasize the $Z=2$ and $Z=1$ crossover caused by the crystal non-uniformity.}
		\label{figure_16}
	\end{center}
\end{figure}

Results of the energy-light calibration for hydrogen and helium isotopes are summarized in Figure~\ref{figure_16}, using the same color scheme adopted for the above discussion as a comparison. The low energy region delimited by the black box in the figure is expanded in the inset for clarity. The light output non-linearity observed for lighter isotopes results in a crossover of the $Z=1$ and $Z=2$ lines mainly caused by the sensitivity of the $Z=1$ light output to the spatial non-uniformity of the scintillation efficiency, as already pointed out while discussing Figure~\ref{figure_08} and easily visible in the zoom shown by the inset of Figure~\ref{figure_16}. These effects don’t play a considerable role for $Z\geq2$ isotopes in the energy range constrained by this investigation. However, a sensitivity to crystal non-uniformities could be observed in heavier isotopes when the energies are sufficiently large for the ion to have a macroscopic (of the order of few cm) range in the crystal. The absence of firm energy-light constraints in the energy regions of $Z\geq2$ affected by the crystal non-linearity makes it extremely difficult to perform a systematic correction of the resulting effects. This is a limitation, as also pointed out in \cite{Wag01}, for the energy resolution of the crystal in applications involving the detection of high-energy particles. Experiments involving direct beams with different energies to directly strike the crystal at different depths, could help in performing a systematic study of non-uniformity effects experienced by heavier ions.
\section{Conclusions}
In this paper, we have studied the light response to light ions of long ($\approx10$ cm) CsI(Tl) crystals with a photodiode readout. In real experimental applications, these crystals are assembled in an array of $4$ closely packed crystals and used as the second detection stage of the HiRA10 array. High quality data collected in a HiRA10 NSCL experimental campaign, combined with data obtained using direct low-energy beams delivered by the tandem accelerator of WMU, allowed us to effectively constrain the light output in a wide energy range for hydrogen (from $1$ MeV up to energies of around $200$ MeV of $^1$H and $300$ MeV of $^3$H) and $2\leq Z\leq4$  (for energies up to $250$ MeV) isotopes. The spatial non-uniformity of the crystals was probed by using a $^{241}$Am $\alpha$-particle source to scan several points of the crystal over the entrance and back surface. A $^{137}$Cs $\gamma$-source was used to longitudinally scan the crystal, indicating linearly decreasing, depth-dependent light efficiency. Based on this assumption, we develop a simplified model to study gradient effects on the light output produced by several incident ions and mainly arising from the combination of the spatial non-uniformity of Tl activator concentration and the geometrical light collection efficiency. Because of the longer range experienced by hydrogen isotopes in the CsI, the $Z=1$ light-output calibration results are significantly affected by the spatial non-uniformity of the crystal, while the crystal appears almost uniform to heavier ($Z>1$) isotopes in the energy domain used to constrain the light output. Accounting for the functional light output derived by our model, we propose a new empirical formula for hydrogen light-output calibration in spatially non-uniform crystals of the type studied here. The formula contains an explicit dependence on the ion mass and allowed us to obtain a simultaneous fit of hydrogen energy-light data.
CsI crystal non-linearities are found to play a significant role in the energy-light calibration of light particles. A proper energy-light calibration across a wide dynamic range requires corrections for the deriving gradient effects, which represent the principal limitation to the CsI(Tl) energy resolution in experiments involving a wide range of energetic light particles. Experiments with light beams directly impinging on the crystal at different depths are therefore required to investigate and correct for non-linearity effects affecting the high-energy region of the energy-light calibration, and typically not easily observable within the energy constraints used to calibrate the response of CsI to $Z>1$ isotopes.
\section*{Acknowledgements}
This work is supported by the US National Science Foundation Grant No. PHY-1565546, U.S. Department of Energy (Office of Science) under Grant Nos. DE-SC0014530, DE-NA0002923. The authors would like to thank the HiRA collaboration for the use of their data in this paper. We also acknowledge the staff of the National Superconducting Cyclotron Laboratory and of the Western Michigan University (Department of Physics) for delivering the high-quality beams used in the present analysis and for assisting us during the preparation of the experiments. We thank Prof. Giacomo Poggi (Firenze) for useful discussions.

\bibliography{DellAquila_CsICalibrations}

\begin{thebibliography}{10}
\expandafter\ifx\csname url\endcsname\relax
  \def\url#1{\texttt{#1}}\fi
\expandafter\ifx\csname urlprefix\endcsname\relax\def\urlprefix{URL }\fi
\expandafter\ifx\csname href\endcsname\relax
  \def\href#1#2{#2} \def\path#1{#1}\fi

\bibitem{Meijer87}
R.~Meijer, et~al., Nucl. Instr. Meth. Phys. Res. A 256 (1987) 521.
\newblock \href {http://dx.doi.org/10.1016/0168-9002(87)90296-8}
  {\path{doi:10.1016/0168-9002(87)90296-8}}.

\bibitem{LeNeindre02}
N.~L. Neindre, et~al., Nucl. Instr. Meth. Phys. Res. A 490 (2002) 251.
\newblock \href {http://dx.doi.org/10.1016/S0168-9002(02)01008-2}
  {\path{doi:10.1016/S0168-9002(02)01008-2}}.

\bibitem{Pout95}
J.~Pouthas, et~al., Nucl. Instr. Meth. Phys. Res. A 357 (1995) 418.
\newblock \href {http://dx.doi.org/10.1016/0168-9002(94)01543-0}
  {\path{doi:10.1016/0168-9002(94)01543-0}}.

\bibitem{Russ15}
P.~Russotto, et~al., Phys. Rev. C 91 (2015) 014610.
\newblock \href {http://dx.doi.org/10.1103/PhysRevC.91.014610}
  {\path{doi:10.1103/PhysRevC.91.014610}}.

\bibitem{Lomb10}
I.~Lombardo, et~al., Nucl. Phys. A 834 (2010) 458c.
\newblock \href {http://dx.doi.org/10.1016/j.nuclphysa.2010.01.063}
  {\path{doi:10.1016/j.nuclphysa.2010.01.063}}.

\bibitem{Wuen09}
S.~Wuenschel, et~al., Nucl. Instr. Meth. Phys. Res. A 604 (2009) 578.
\newblock \href {http://dx.doi.org/10.1016/j.nima.2009.03.187}
  {\path{doi:10.1016/j.nima.2009.03.187}}.

\bibitem{Mor16}
L.~Morelli, et~al., J. Phys. G: Nucl. Part. Phys. 43 (2016) 045110.
\newblock \href {http://dx.doi.org/10.1088/0954-3899/43/4/045110}
  {\path{doi:10.1088/0954-3899/43/4/045110}}.

\bibitem{Sal16}
F.~Salomon, et~al., J. Instr. 11 (2016) C01064.
\newblock \href {http://dx.doi.org/10.1088/1748-0221/11/01/C01064}
  {\path{doi:10.1088/1748-0221/11/01/C01064}}.

\bibitem{Past17}
G.~Pastore, et~al., Nucl. Instr. Meth. Phys. Res. A 860 (2017) 42.
\newblock \href {http://dx.doi.org/10.1016/j.nima.2017.01.048}
  {\path{doi:10.1016/j.nima.2017.01.048}}.

\bibitem{Davin01}
B.~Davin, et~al., Nucl. Instr. Meth. Phys. Res. A 473 (2001) 302.
\newblock \href {http://dx.doi.org/10.1016/S0168-9002(01)00295-9}
  {\path{doi:10.1016/S0168-9002(01)00295-9}}.

\bibitem{Poll05}
E.~Pollacco, et~al., Eur. Phys. J. A 25 (2005) 287.
\newblock \href {http://dx.doi.org/10.1140/epjad/i2005-06-162-5}
  {\path{doi:10.1140/epjad/i2005-06-162-5}}.

\bibitem{Acosta12}
L.~Acosta, et~al., EPJ Web. Conf. 31 (2012) 00035.
\newblock \href {http://dx.doi.org/10.1051/epjconf/20123100035}
  {\path{doi:10.1051/epjconf/20123100035}}.

\bibitem{Pag16}
E.~Pagano, et~al., EPJ Web. Conf. 117 (2016) 10008.
\newblock \href {http://dx.doi.org/10.1051/epjconf/201611710008}
  {\path{doi:10.1051/epjconf/201611710008}}.

\bibitem{Wall07}
M.~Wallace, et~al., Nucl. Instr. Meth. Phys. Res. A 583 (2007) 302.
\newblock \href {http://dx.doi.org/10.1016/j.nima.2007.08.248}
  {\path{doi:10.1016/j.nima.2007.08.248}}.

\bibitem{Dell18}
D.~Dell'Aquila, et~al., Nucl. Instr. Meth. Phys. Res. A 877 (2018) 227.
\newblock \href {http://dx.doi.org/10.1016/j.nima.2017.09.046}
  {\path{doi:10.1016/j.nima.2017.09.046}}.

\bibitem{Par02a}
M.~P$\hat{a}$rlog, et~al., Nucl. Instr. Meth. Phys. Res. A 482 (2002) 674.
\newblock \href {http://dx.doi.org/10.1016/S0168-9002(01)01710-7}
  {\path{doi:10.1016/S0168-9002(01)01710-7}}.

\bibitem{Birks64}
J.~Birks, The Theory and Practice of Scintillation Counting, Pergamon Press,
  Oxford, 1964.

\bibitem{Meyer62}
A.~Meyer, R.~Murray, Phys. Rev. 128 (1962) 98.
\newblock \href {http://dx.doi.org/10.1103/PhysRev.128.98}
  {\path{doi:10.1103/PhysRev.128.98}}.

\bibitem{Par02b}
M.~P$\hat{a}$rlog, et~al., Nucl. Instr. Meth. Phys. Res. A 482 (2002) 693.
\newblock \href {http://dx.doi.org/10.1016/S0168-9002(01)01712-0}
  {\path{doi:10.1016/S0168-9002(01)01712-0}}.

\bibitem{Lopez18}
O.~Lopez, et~al., Nucl. Instr. Meth. Phys. Res. A 884 (2018) 140.
\newblock \href {http://dx.doi.org/10.1016/j.nima.2017.12.041}
  {\path{doi:10.1016/j.nima.2017.12.041}}.

\bibitem{Horn92}
D.~Horn, et~al., Nucl. Instr. Meth. Phys. Res. A 320 (1992) 273.
\newblock \href {http://dx.doi.org/10.1016/0168-9002(92)90785-3}
  {\path{doi:10.1016/0168-9002(92)90785-3}}.

\bibitem{Laroc94}
Y.~Larochelle, et~al., Nucl. Instr. Meth. Phys. Res. A 348 (1994) 167.
\newblock \href {http://dx.doi.org/10.1016/0168-9002(94)90856-7}
  {\path{doi:10.1016/0168-9002(94)90856-7}}.

\bibitem{Twen90}
C.~Twenh$\ddot{o}$fel, et~al., Nucl. Instr. Meth. Phys. Res. A 51 (1990) 58.
\newblock \href {http://dx.doi.org/10.1016/0168-583X(90)90539-7}
  {\path{doi:10.1016/0168-583X(90)90539-7}}.

\bibitem{Aiello96}
S.~Aiello, et~al., Nucl. Instr. Meth. Phys. Res. A 369 (1996) 50.
\newblock \href {http://dx.doi.org/10.1016/0168-9002(95)00763-6}
  {\path{doi:10.1016/0168-9002(95)00763-6}}.

\bibitem{Zieg85}
J.~Ziegler, J.~Biersack, U.~Littmark, The Stopping and Range of Ions in Matter,
  Pergamon Press, New York, 1985.

\bibitem{Scionix}
SCIONIX Dedicated Scintillation Detectors, Regulierenring 5, 3981 LA Bunnik,
  The Netherlands.

\bibitem{Man18}
J.~Manfredi, et~al., Nucl. Instr. Meth. Phys. Res. A 888 (2018) 177.
\newblock \href {http://dx.doi.org/10.1016/j.nima.2017.12.082}
  {\path{doi:10.1016/j.nima.2017.12.082}}.

\bibitem{RomerArm}
Hexagon Manufacturing Intelligence, ROMER Absolute Arm -- Coordinate Measuring
  Machine.

\bibitem{Wag01}
A.~Wagner, et~al., Nucl. Instr. Meth. Phys. Res. A 456 (2001) 290.
\newblock \href {http://dx.doi.org/10.1016/S0168-9002(00)00542-8}
  {\path{doi:10.1016/S0168-9002(00)00542-8}}.

\bibitem{Gong88}
W.~Gong, et~al., Nucl. Instr. Meth. Phys. Res. A 268 (1988) 190.
\newblock \href {http://dx.doi.org/10.1016/0168-9002(88)90605-5}
  {\path{doi:10.1016/0168-9002(88)90605-5}}.

\bibitem{gal95}
J.~G\'al, et~al., Nucl. Instr. Meth. Phys. Res. A 366 (1995) 120.
\newblock \href {http://dx.doi.org/10.1016/0168-9002(95)00560-9}
  {\path{doi:10.1016/0168-9002(95)00560-9}}.

\end{thebibliography}

\end{document}